\documentclass[iop]{emulateapj}
\usepackage{amsmath,amssymb}
\usepackage[colorlinks,citecolor=blue,linkcolor=blue,urlcolor=black]{hyperref}
\usepackage{etoolbox}

\makeatletter

\patchcmd{\NAT@citex}
  {\@citea\NAT@hyper@{%
     \NAT@nmfmt{\NAT@nm}%
     \hyper@natlinkbreak{\NAT@aysep\NAT@spacechar}{\@citeb\@extra@b@citeb}%
     \NAT@date}}
  {\@citea\NAT@nmfmt{\NAT@nm}%
   \NAT@aysep\NAT@spacechar\NAT@hyper@{\NAT@date}}{}{}

\patchcmd{\NAT@citex}
  {\@citea\NAT@hyper@{%
     \NAT@nmfmt{\NAT@nm}%
     \hyper@natlinkbreak{\NAT@spacechar\NAT@@open\if*#1*\else#1\NAT@spacechar\fi}%
       {\@citeb\@extra@b@citeb}%
     \NAT@date}}
  {\@citea\NAT@nmfmt{\NAT@nm}%
   \NAT@spacechar\NAT@@open\if*#1*\else#1\NAT@spacechar\fi\NAT@hyper@{\NAT@date}}
  {}{}

\begin{document}               

\title{High Metallicity LGRB Hosts}
\author{J. F. Graham$^{\hyperref[1]{1}}$ \affil{Max-Planck Institute for Extraterrestrial Physics, Giessenbachstrasse 1, 85748 Garching, Germany}}
\author{A. S. Fruchter$^{\hyperref[2]{2}}$\affil{Space Telescope Science Institute, 3700 San Martin Drive, Baltimore MD 21218}}
\author{E. M. Levesque$^{\hyperref[3]{3}}$\affil{University of Colorado at Boulder, UCB 391, Boulder, CO 80309}}
\author{L. J. Kewley$^{\hyperref[4]{4}}$\affil{Australian National University, Research School of Astronomy \& Astrophysics, Mount Stromlo Observatory, Cotter Road, Weston Creek, ACT 2611, Australia}}
\author{N. R. Tanvir$^{\hyperref[5]{5}}$\affil{Department of Physics and Astronomy, University of Leicester, University Road, Leicester LE1 7RH}}
\author{A. J. Levan$^{\hyperref[6]{6}}$\affil{Department of Physics, University of Warwick, Coventry CV4 7AL}}
\author{S. K. Patel$^{\hyperref[7]{7}}$\affil{National Space Science \& Technology Center, Universities Space Research Administration, 320 Sparkman Drive, Huntsville AL 35805}}
\author{K. Misra$^{\hyperref[8]{8}}$\affil{Aryabhatta Research Institute of Observational Sciences (ARIES), Manora Peak, Nainital-263002, Uttarakhand, India}}
\author{K.-H. Huang$^{\hyperref[9]{9}}$\affil{University of California Davis, 1 Shields Avenue, Davis, CA, 95616}} 
\author{D. E. Reichart$^{\hyperref[10]{10}}$\affil{Department of Physics and Astronomy, University of North Carolina at Chapel Hill, Campus Box 3255, Chapel Hill, NC 27599}}
\author{M. Nysewander$^{\hyperref[10]{10}}$\affil{Department of Physics and Astronomy, University of North Carolina at Chapel Hill, Campus Box 3255, Chapel Hill, NC 27599}}
\author{P. Schady$^{\hyperref[1]{1}}$ \affil{Max-Planck Institute for Extraterrestrial Physics, Giessenbachstrasse 1, 85748 Garching, Germany}}
\affil{$^1$\label{1} Max-Planck Institute for Extraterrestrial Physics, Giessenbachstrasse 1, 85748 Garching, Germany\\
$^2$\label{2} Space Telescope Science Institute, 3700 San Martin Drive, Baltimore MD 21218\\
$^3$\label{3} University of Colorado at Boulder, UCB 391, Boulder, CO 80309\\
$^4$\label{4} Australian National University, Mount Stromlo Observatory, Cotter Road, Weston Creek, ACT 2611, Australia\\
$^5$\label{5} Department of Physics and Astronomy, University of Leicester, University Road, Leicester LE1 7RH\\
$^6$\label{6} Department of Physics, University of Warwick, Coventry CV4 7AL\\
$^7$\label{7} National Space Science \& Technology Center, 320 Sparkman Drive, Huntsville AL 35805\\
$^8$\label{8} Aryabhatta Research Institute of Observational Sciences (ARIES), Manora Peak, Nainital-263002, Uttarakhand, India\\
$^9$\label{9} University of California Davis, 1 Shields Avenue, Davis, CA, 95616\\
$^{10}$\label{10} Department of Physics and Astronomy, University of North Carolina at Chapel Hill, Chapel Hill, NC 27599}

\journalinfo{}
\submitted{}


\begin{abstract}

We present our imaging and spectroscopic observations of the host galaxies of two dark long bursts with anomalously high metallicities, LGRB 051022 and LGRB 020819B, which in conjunction with another LGRB event with an optical afterglow \citep{Levesque2} comprise the three LGRBs with high metallicity host galaxies in the \cite{stats_paper} sample.  In \cite{stats_paper}, we showed that LGRBs exhibit a strong and apparently intrinsic preference for low metallicity environments (12+log(O/H) $<$ 8.4 in the KK04 scale) in spite of these three cases with abundances of about solar and above.  Not only do these exceptions not share the typical low metallicities of LGRB hosts, they are consistent with the general star-forming galaxy population of comparable brightness \& redshift.  This result is intrinsically surprising: even among a preselected sample of the high metallicity LGRBs, were the metal aversion to remain in effect for these objects, we would expect the metallicity these LGRBs to still be lower than the typical metallicity for the galaxies at that luminosity and redshift (i.e., either a outlier of said population, or among the lowest galaxies available within it). Therefore we deduce that it is likely possible to form an LGRB in a high metallicity environment although with greater rarity.

From this we conclude that there are three possible explanations for the presence of the LGRBs observed in high metallicity hosts as seen to date:  (1) LGRBs do not occur in high metallicity environments and those seen in high metallicity hosts are in fact occurring in low metallicity environments that have become associated with otherwise high metallicity hosts but remain unenriched. (2) The LGRB formation mechanism while preferring low metallicity environments does not strictly require it resulting in a gradual decline in burst formation with increasing metallicity. (3) The typical low metallicity LGRBs and the few high metallicity cases are the result of physically different burst formation pathways with only the former affected by the metallicity and the later occurring much more infrequently. 

\end{abstract}

\section{Introduction}

Long soft gamma-ray bursts (LGRBs) are frequently found in a particular type of host galaxy: blue irregulars \citep{Fruchter1999, Fruchter, LeFlochblue, LeFlochblue2002}.  LGRBs have shown a strong preference for occurring in star-forming galaxies \citep{Fruchter1999, Christensen, LeFloch2006}, which often exhibit bright emission lines \citep{Bloom, Vreeswijk, Levesque051022} indicative of substantial populations of young, massive stars.  LGRBs have also frequently been associated with broad-lined Type Ic (Ic-bl) supernovae (SNe) \citep{Stanek2003, Hjorth2003, Woosley} so named because their spectral lines show broadening from high-velocity ($\sim$15,000 km s$^{-1}$) ejecta.

\cite{Fruchter} performed a detailed study of the LGRB host galaxy population, using the GOODS core-collapse supernovae sample as a comparative group, and shows a surprising bias against hosts being grand design spiral galaxies. Only one out of 42 LGRB host galaxies was a grand design spiral (as compared to the GOODS supernovae sample where about half the host galaxies were spirals).  If one constrains the LGRB host population to a redshift of 1.2 or less so as to match the redshift distribution of the supernovae sample this drops to one out of 18, still a rather surprising result.  The remainder of the GRB host population are composed of generally faint, irregular galaxies, whereas by contrast, half of all the core collapse supernovae hosts were spirals.  Additionally, the \cite{Fruchter} sample showed a strong preference for LGRB's occurring in the brightest, and hence likely the most star-forming regions of their hosts.

One of the conclusions of \cite{Fruchter} was that LGRB formation requires a low metallicity progenitor and the bias toward irregular galaxies is a result of their low metallicity as expected by the mass-metallicity relation.  This conclusion was supported by subsequent works of Stanek et al.\thinspace (2006), where a comparison of LGRB hosts and galaxies in the Sloan sample (of similar magnitude) shows that the very nearest hosts have low metallicity.  Most convincingly, \cite{Modjaz2008} compared the host metallicity luminosity relation of LGRBs with known Ic supernovae counterparts to nearby Ic Supernovae without LGRB associations (see Figure 5 in \citealt{Modjaz2008}).  They found a dramatic difference in metallicity between the two samples, even when host luminosity is accounted for, showing a profound metallicity avoidance for the LGRBs.  \cite{Kocevski} compared the mass distribution of LGRB hosts to the general star-forming galaxy population as a function of redshift, finding that the populations remain dissimilar out to z $\sim$ 1 with an upper limit on the stellar mass of LGRB hosts evolving with redshift. This favors a smooth decrease in the LGRB formation rate with increasing metallicity above 12+log(O/H) = 8.7 over an extremely low metallicity cut-off thus limiting LGRB hosts to low mass spirals and dwarfs at low redshift and suggesting that LGRBs remain metallicity biased tracers of star-formation out to intermediate redshifts.  On the other hand \cite{Savaglio} argue that LGRB hosts lie on the same mass-metallicity relation as regular galaxies.  \cite{Berger2006} have used this claim to argue that because the host of GRB 020127 is unusually bright it must also be metal-rich.  \cite{Peeples} however show the existence of low metallicity outliers on the luminosity-metallicity relation, and argue against the assignment of metallicities (to individual galaxies) based only on their luminosities.  In particular they highlight the morphologically similarities of their bright outliers to the brighter hosts in the \cite{Fruchter} sample.  \cite{MannucciLGRBs} claimed that the observed low metallicity LGRB bias is not an intrinsic preference but only a byproduct of the higher star-formation rates observed in LGRB hosts and the \cite{Mannucci} anti-correlation between mass, star-formation rate, and metallicity.

In \cite{stats_paper}, we showed that LGRBs exhibit a strong and apparently intrinsic preference for environments with metallicities below 12+log(O/H) $<$ 8.4 on the KK04 \citep{KobulnickyKewley} scale.  However we note therein that some exceptions do exist to this trend --- three of the 14 LGRB in the sample possess abundances of about solar and above.  While the majority of the LGRB population is constrained to low metallicities of about a third solar and below these exceptions probably show that is it still possible to form an LGRB in a high metallicity environment although with greater rarity.  For us to use LGRBs to trace the star-formation of the Universe (even at low metallicities), we must understand the conditions required for their production and thus selection effects that take place even before we see the LGRB.  The implications of these high metallicity bursts are important not only for understanding the formation of LGRBs but also for any hope of being ever able to use them as cosmological probes. 

Here, we specifically address these exceptions by beginning with a more detailed examination of the individual cases, LGRBs 051022, 020819B, and 050826, followed by an analysis of these objects as a population. Initial metallicity results for the LGRB 051022 and 020819B host galaxies were published in our conference proceedings \citep{conference_proceedings} and a short {\apjl} letter \citep{Levesque020819B} respectively and we perform a more detailed metallicity analysis (particularly with regard to error), study additional object parameters, and give greater consideration of the scientific impact of these results herein.  We will also briefly discuss LGRB 050826, particularly with regard to its host properties.   Both LGRB 051022 and 020819B are among the brightest LGRB host galaxies yet seen (L* or above in the Schechter luminosity function), and via the luminosity-mass-metallicity relationship, such galaxies would be expected to possess a comparatively higher metallicity than that yet seen in long burst hosts.  Both are at a low enough redshift that the emission lines necessary to study their metallicity are visible in either the optical or near-IR.  Also both lack an optical transient or SNe contribution, classifying them as
dark bursts (See \cite{Roldark, Jakobssondark} for more detail on dark bursts and their criteria for classification).

\cite{Perley2013} performed a detailed study of 23 dark bursts and identified their host galaxies to be more massive, more star-forming, and more dust obscured than LGRB hosts with optical afterglows.  While this likely implies that dark bursts also are more metal rich than their optically visible counterparts both \cite{stats_paper} and \cite{Perley2013} conclude that, even after accounting for the contribution of dark bursts, a strong preference for LGRB to occur in low metallicity host galaxies remains.  Furthermore \cite{Perley_and_Perley} suggest that per unit underlying star-formation the outer parts of submillimeter galaxies overproduce LGRBs while the inner parts underproduce them consistent with a metallicity or IMF gradient therein.


Our third object however, LGRB 050826, did have an optical transient establishing that such bursts do also exist in high metallicity environments.  This furthers the argument that dark bursts are just normal LGRBs with heavy dust extinction attenuating the viable afterglow.  LGRB 050826 was also the first case where the a high metallicity LGRB was found without without previous suspicion that this was likely the case before obtaining the metallicity measurement.  After examining these cases individually and as a population, we will propose and discuss explanations for the existence of high metallicity LGRBs in light of the now established intrinsic preference of LGRBs for low metallicity environments.  Such an explanation must reconcile a formation mechanism that is biased against a metal rich environment with one that seems to still permit some formation against that bias.


\section{Observations}

\subsection{GRB 051022} 
\subsubsection{Burst Detection and Localization}

LGRB 051022 was initially detected by all three instruments on the High Energy Transient Explorer 2 (HETE-2) satellite, the FREnch GAmma TElescope (Fregate), Wide-field X-ray Monitor (WXM), and the Soft X-ray Camera (SXC) (\citealt{GCN4131} GCN 4131).  The burst was sufficiently bright in soft X-rays that the SXC position was determined independently of the WXM location (\citealt{GCN4137} GCN 4137).  HETE-2 observations indicated a burst duration in excess of 2 to 4 minutes securely identifying this as a long gamma ray burst  (\citealt{GCN4137} GCN 4137).  Within the HETE-2 error circles a fading afterglow was detected in the X-ray, mm, and radio bands.  In spite of this, no variable optical source was found at the afterglow position (\citealt{Rol051022}, GCN 4134, GCN 4143) thus placing this LGRB in a special class of bursts called the ``dark'' bursts.  The afterglow position identifies a host galaxy at a redshift of z = 0.8 (\citealt{GCN4156} GCN 4156).  See Table \ref{detection} for details of prompt and afterglow localizations.

\begin{table*}[ht]
\begin{center}
\caption{\label{detection}}
\resizebox{\textwidth}{!}{
\vspace{-0.1 cm}
\begin{tabular}{cccccccccccccccc}
\hline
\noalign{\vskip 0.5mm} 
\hline
        $\Delta$T (days) & RA    &   Dec  & Error & Instrument & Band & Reference \\
\hline
0      & 23$^{h}$55$^{m}$53$^{s}$ & +19$^{\circ}$37$^{m}$43$^{s}$     & 14'  & HETE-2 WXM & 2-25 keV & \citealt{GCN4131} GCN 4131\\
0      & 23$^{h}$56$^{m}$00$^{s}$ & +19$^{\circ}$35$^{m}$51$^{s}$     & 2.5' & HETE-2 SXC & 0.5-10 keV & \citealt{GCN4131} GCN 4131; \citealt{GCN4137} GCN 4137\\
0.15 & 23$^{h}$56$^{m}$04.1$^{s}$ & +19$^{\circ}$36$^{m}$25.1$^{s}$ & 4" & \emph{Swift} XRT & X-ray & \cite{GCN4141} GCN 4141\\
1.35 & 23$^{h}$56$^{m}$04.1$^{s}$ & +19$^{\circ}$36$^{m}$25.4$^{s}$ & 0.5" & Plateau de Bure Interferometer & 3 mm & \cite{GCN4157} GCN 4157, \cite{CastroTirado}\\
1.54 & 23$^{h}$56$^{m}$04.1$^{s}$ & +19$^{\circ}$36$^{m}$24.1$^{s}$ & 1" & VLA & 8.4, 4.9 and 1.4GHz & \cite{GCN4154} GCN 4154\\
3.46 & 23$^{h}$56$^{m}$04.1$^{s}$ & +19$^{\circ}$36$^{m}$23.9$^{s}$ & 0.7" & {\it Chandra} ACIS-S & X-ray & \cite{GCN4163} GCN 4163, \cite{Rol051022} \\
\hline
\end{tabular}}
\end{center}
\vspace{-0.2 cm}
Prompt and afterglow localizations of LGRB 051022.
\end{table*}


\subsubsection{Burst Host Imaging} \label{051022host}

\begin{figure}[ht]
\begin{center}
\includegraphics[width=.48\textwidth]{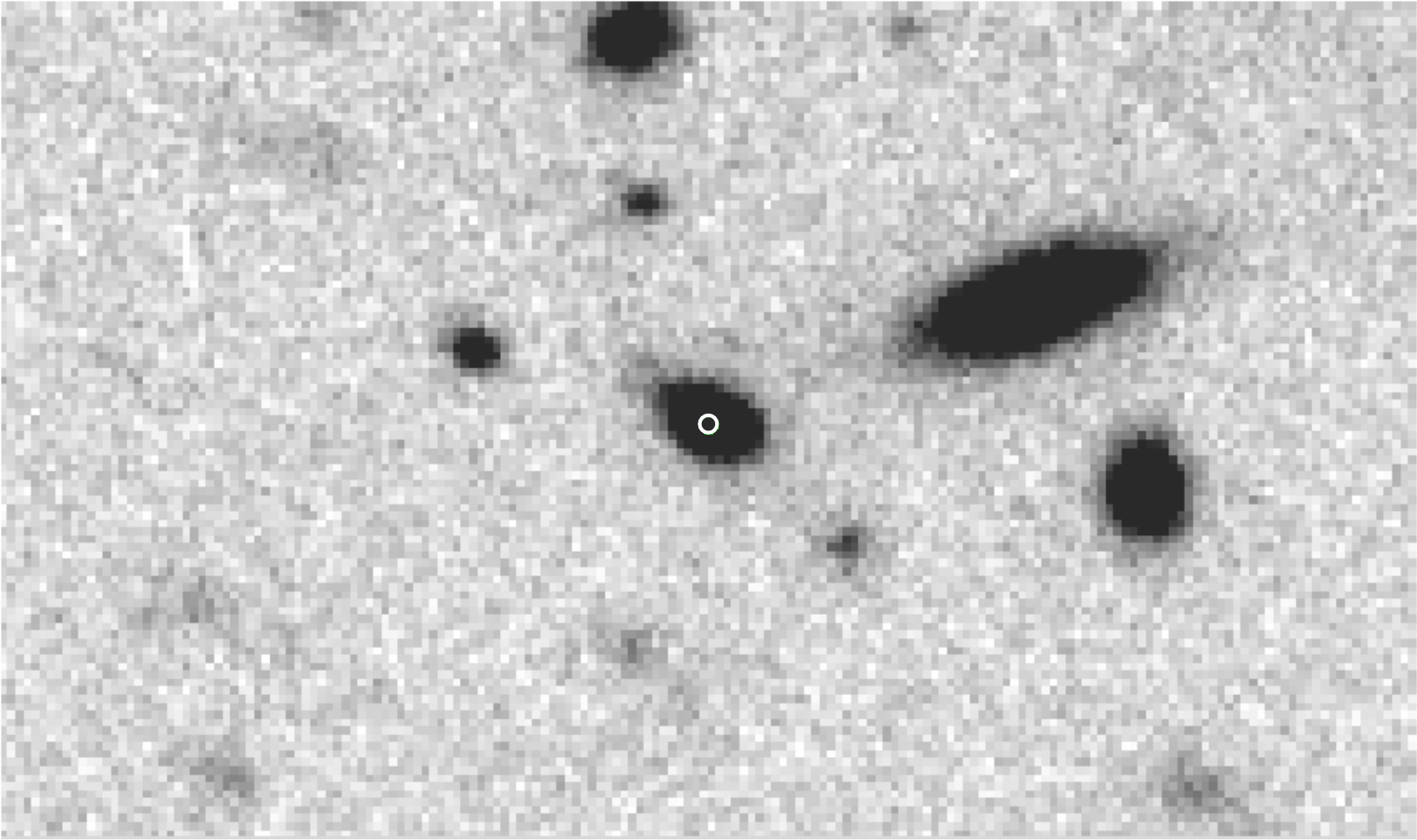}
\caption{GRB051022 host galaxy (\textit{r} band) with our astrometrically matched 0.17" 1 sigma radius Chandra X-ray error circle overplotted.  Note that the morphology of this host is unresolved in ground-based images.   \label{image}}
\end{center}
\end{figure}

We obtained imaging on the LGRB 051022 field in \textit{r}, \textit{i}, and \textit{z} bands with the GMOS instrument on Gemini North on October 22nd, 2005. A later image, for image subtraction, was acquired on October 25th, 2005. More details on observations and data analysis are presented in \cite{Rol051022}. The resultant host magnitudes in \textit{r}, \textit{i}, and \textit{z} bands are 22.04 $\pm$ 0.01, 21.77 $\pm$ 0.01, and 21.30 $\pm$ 0.04  respectively (\citealt{Rol051022}).  The \textit{r} band image is shown in figure \ref{image} with the astrometric solution determined in section \ref{astrometry}.

Approximating the redshift as a straight \textit{i} to \textit{B} band central wavelength conversion yields a shift in central wavelength from 780 nm (for the \textit{i} band filter) to 432 nm {vs.} 438 nm for a \textit{B} band filter.  Preserving the flux values gives a rest frame absolute \textit{B} band magnitude of -21.60 $\pm$ 0.01 for the host galaxy of GRB 051022.  The estimated absolute magnitude is unusually bright for an LGRB host and belongs at about 1.4 $\times$ L* on the Schechter luminosity function (we adopt blue galaxy M$^{*}_{B}$ values from Table 3 of  \citealt{L*} for all L* comparisons and as is typical for L* we do not attempt to correct for extinction).  

\begin{figure*}[ht]
\begin{center}
\includegraphics[width=.48\textwidth]{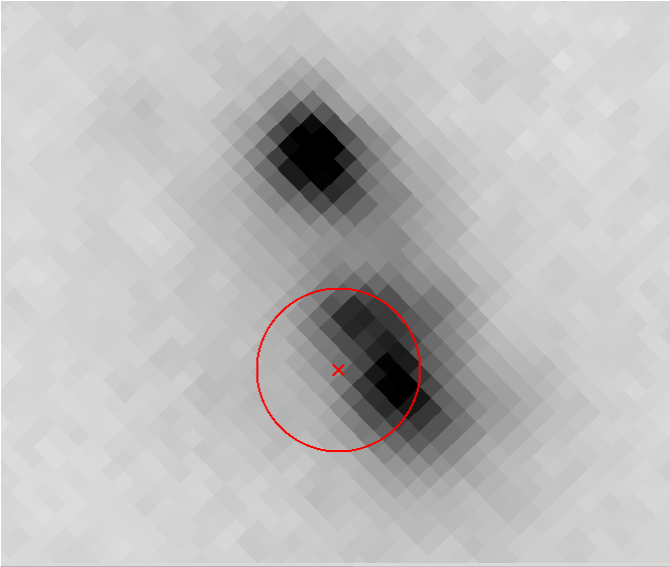}
\includegraphics[width=.48\textwidth]{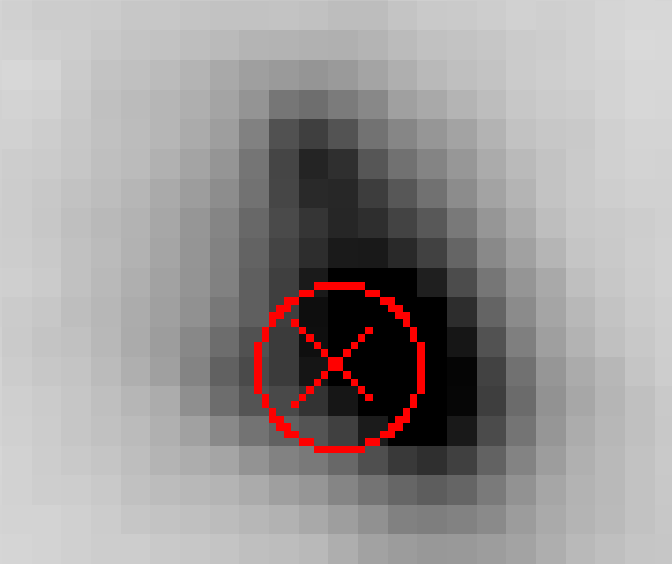}
\caption{ACS/WFC F606W (left) and WFC3/IR F160W (right) images of the LGRB 051022 host galaxy.  The images were astrometrically aligned to our Gemini GMOS imaging and the astrometric fit from section \ref{astrometry} was transformed to the HST images giving the 0.17" 1 sigma radius Chandra X-ray error circle shown.  The images are also shown aligned with each other in the figure.\label{hst_initial}}
\end{center}
\end{figure*}

To discern the host morphology, the field of GRB 051022 was imaged with the Hubble Space Telescope (HST) using repaired ACS/WFC in F606W on August 21$^{st}$ 2009 and one orbit with WFC3/IR in F160W on October 12$^{th}$ 2009 (see figure \ref{hst_initial} left and right respectively).  Four dithered 520 second exposures were obtained in the F606W filter of the ACS WFC, and four dithered 600 second exposures were obtained in the F160W filter of the WFC3 IR channel.  The two data sets were separately combined using the drizzle/multidrizzle packages and process outlined in \cite{drizzle, koekemoer}.  To allow a localized color comparison the combined ACS field was also mapped to the scale of the WFC image using ``blot'' \citep{drizzle}.  Using a consistent aperture across both bands (for optimal color term accuracy) we determine a host magnitude of 21.903 in F606W and 20.608 in F160W relative to the instrument zero points.  Adopting the noise correlation ratio of \cite{drizzle} equation 9 we calculate a statistical error of $\pm$ 0.006 magnitudes on both measurements. For absolute photometry we estimate out of aperture flux lost at 0.05 mags or less.

\begin{figure*}[ht]
\begin{center}
\includegraphics[width=.325\textwidth]{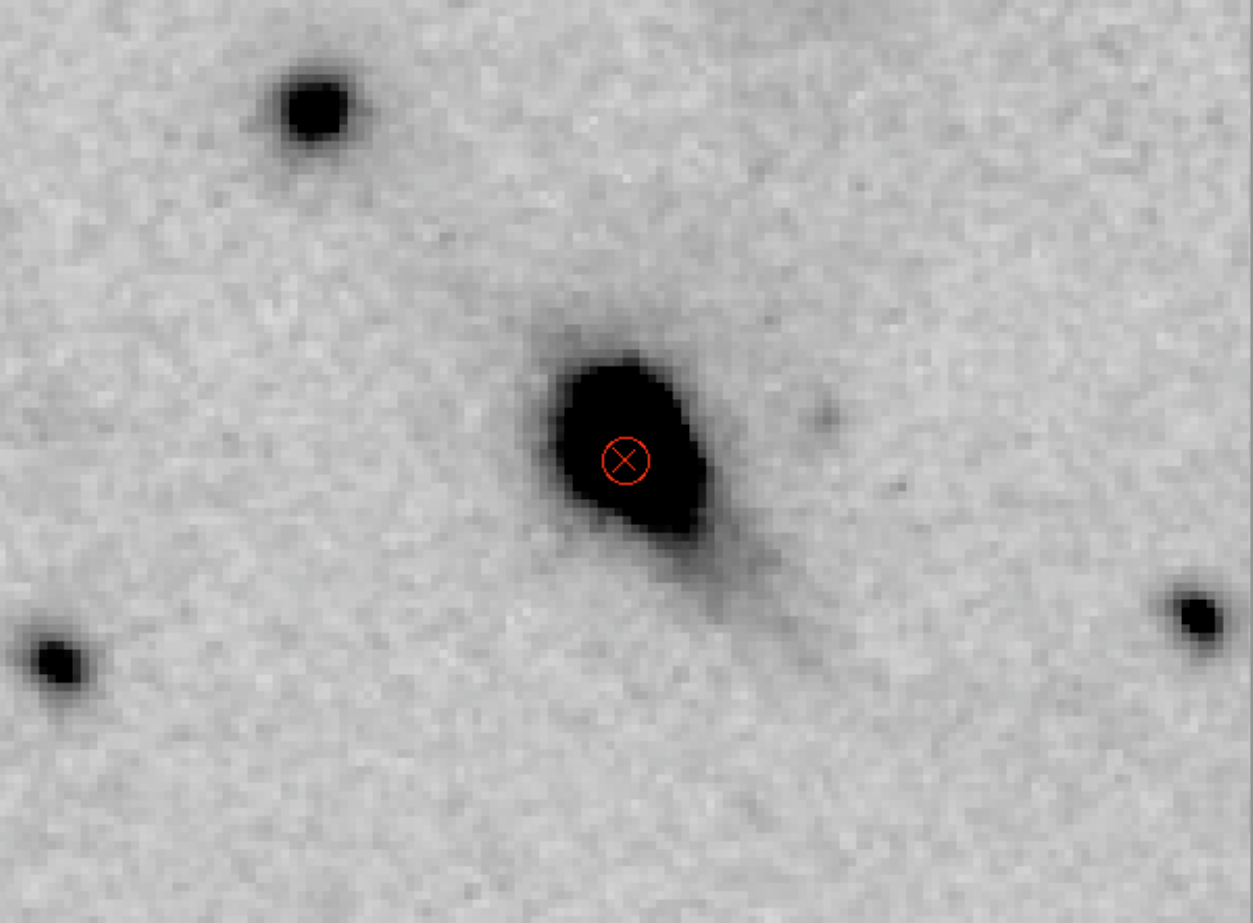}
\includegraphics[width=.325\textwidth]{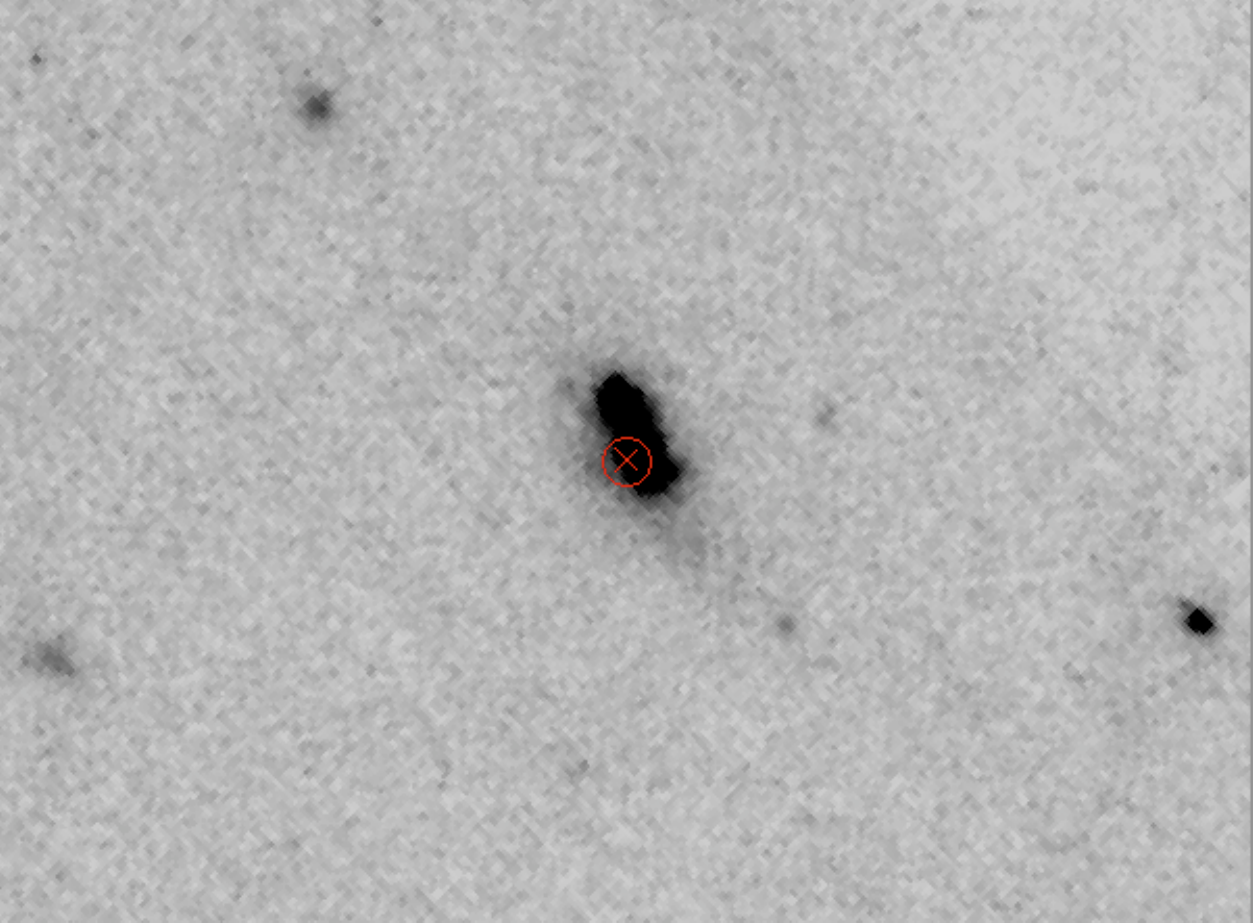}
\includegraphics[width=.325\textwidth]{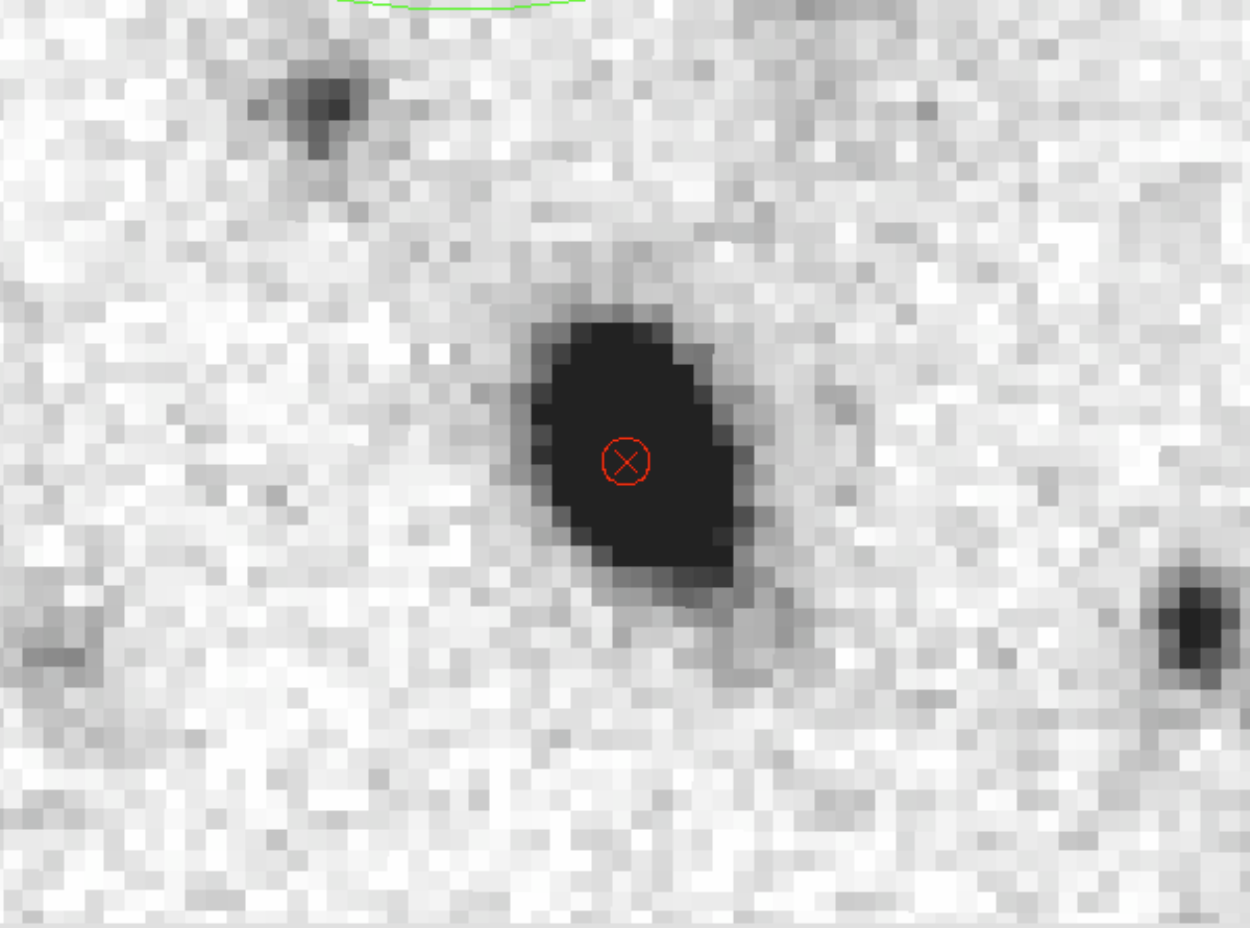}
\caption{Astrometrically matched WFC3/IR F160W (left), ACS/WFC F606W (center), and Gemini GMOS \textit{r} band images of the LGRB 051022 host galaxy showing the tidal tail emanating south west from the host galaxy.  The 0.17" 1 sigma radius Chandra X-ray error circle from from section \ref{astrometry} is shown as before.  The presence of a tidal tail identifies the system as a merger as speculated in \cite{conference_proceedings} and explains the system's high star-formation rate and starburst history. \label{tt_image}}
\end{center}
\end{figure*}

The ACS/WFC F606W image is just blue of u band in the host galaxy rest frame and thus provides a map of the active star-formation and likely LGRB progenitor regions.  
The bimodal (perhaps trimodal) nature of the F606W image and the presence of additional regions in the F160W imaging (approximately equivalent to rest frame z band) suggests a clumpy system with varying degrees of star-formation and potentially variant metallicities.  More interesting from a morphological perspective is the presence of a tidal tail emanating south west from the host galaxy (Figure \ref{tt_image} left).  Our Gemini GMOS \textit{r} band imaging also shows faint evidence of a similar extended emission (Figure \ref{tt_image} right) which is less apparent in the ACS/WFC F606W image (Figure \ref{tt_image} center) due to the latter's lower surface brightness sensitivity.

\subsubsection{Astrometric Refinement of GRB 051022 \label{astrometry}}




We obtained 20 ks of {\it Chandra} ACIS observation of the burst afterglow as described in \cite{Rol051022}.  In addition to the afterglow, our observations detected 6 X-ray background sources.  In order to better constrain the location of the burst we identified counterparts for all 6 sources in our Gemini GMOS \textit{r} band imaging and generated an astrometric solution between the X-ray source fits and their optical counterpart positions.  Note the astrometric solution determined and used here is (aside from the X-ray source fitting) independent of the similar analysis presented in \cite{Rol051022}.  (\textit{r} band observations are typically optimal for astrometry as the objects are brighter than in the bluer bands and the GMOS \textit{i} and redder bands have significant fringing which disrupts object centroiding).  We observe that the residual in the astrometric fit of the background sources is proportional to almost exactly the inverse square root of the number of counts.  Given that the brightest of the background sources is only about half the flux of the observed afterglow the error in the X-ray source fitting is dominated by the inaccuracy in centering the background X-ray objects.  The deviation from fitting the residual as the inverse square root of the number of counts we observe to be bounded by a monotonically increasing function of the distance from the image center.  As the LGRB afterglow is centered on the X-ray image this increasing error with distance from the image center is negligible for the X-ray afterglow source.  Therefore, the error on the object position is dominated by the error in the astrometric solution obtained on the background objects.  Thus we adopt the RMS fit accuracy as the error in the our astrometrically matched source position locating the X-ray afterglow to our GMOS image with a one sigma accuracy of 0.17" (or 1.2 binned 2 $\times$ 2 GMOS pixels).  This places the burst at the center of the host galaxy in the ground based image (See Figure \ref{image}).

Using the coordinates of ten 2MASS catalogue objects as an absolute reference we determined astrometric calibration for our Gemini optical images using the starlink astrom package with an error of 0.057" RMS.  Applied in concert with the Chandra X-ray afterglow to Gemini GMOS astrometric solution from the preceding paragraph, this places GRB 051022 at an absolute astrometric position of RA: 23$^{h}$56$^{m}$04.110$^{s}$ Dec: +19$^{\circ}$36$^{m}$24.03$^{s}$ (J2000) with an accuracy of 0.18" statistical and 0.05" systematic.

Astrometric alignment of the ACS/WFC F606W and WFC3/IR F160W reduced drizzled images to the Gemini GMOS \textit{r} band imaging yielded an accuracy (RMS error) of 0.010 and 0.022 arc-seconds respectively.  The RMS error is essentially insignificant when added in quadrature with the refined {\it Chandra} X-ray afterglow error.  This likely locates the burst to the southern pair of star-forming regions and excludes the brighter northern region beyond the two sigma limit (see figure \ref{hst_initial}).

\subsubsection{Host Optical Spectroscopy}
Initial spectroscopic observations were obtained with the GMOS instrument on Gemini North on 
November 25$^{th}$ 2005 for a total exposure time of 1 hour.  The R400 grating offers a reasonable compromise between spectral resolution (1.37 {\AA}/pixel) and width of coverage (about 4000 {\AA}) giving a spectral range of 5500 to 9500 {\AA} for a central wavelength of 7500 {\AA}.  A 50 {\AA} dither in wavelength was also added to ensure continuous spectral coverage across chip gaps and allow for easy removal of other chip based effects.  Due to the abundance of skylines in the spectral range the Nod \& Shuffle method was used offering a dramatic improvement in sky subtractions over conventional spectroscopy due to its more coincident and technically consistent sampling of object and sky spectra.  A brief introduction to the Nod \& Shuffle process is provided in \cite{070714Bpaper}, for a more detailed description of the Nod \& Shuffle process see \cite{Cuillandre} and \cite{Glazebrook}, and for its use on Gemini see \cite{Glazebrookgemini} and \cite{Abraham}.


In order to determine whether nearby galaxies were in association with the host, Nod \& Shuffle Multi-Object Spectroscopy (MOS) was employed to obtain spectra of several objects with no increase in observing time.  Observations consisted of two 30 min Nod \& Shuffle integrations, each containing two 15 min spectra.

The individual spectroscopic exposures were reduced using the standard IRAF  ``Gemini.GMOS"  Nod \& Shuffle packages.  This is essentially the same as a conventional spectroscopic reduction except the two shuffled images on each exposure are subtracted from each other after bias subtraction and before flat fielding.  Due to the small number of Nod \& Shuffle spectra observed, the spectra were combined in 2d to optimize cosmic ray rejection as described in \cite{070714Bpaper}.  A custom Nod \& Shuffle dark was also used.

\begin{figure*}[ht]
\begin{center}
\includegraphics[width=1\textwidth,height=.3\textwidth]{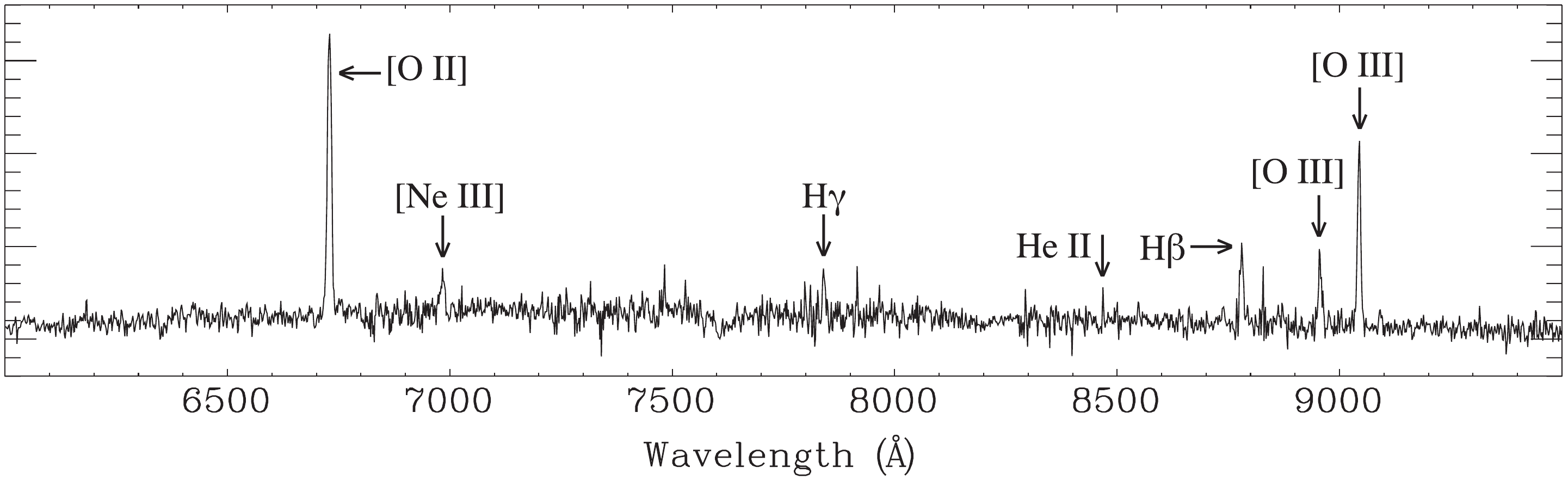}
\caption{Gemini GMOS Nod \& Shuffle optical spectrum of GRB 051022.  The [O II], [O III], and H$\beta$ lines are used for the R$_{23}$ method. \label{spectra}}
\end{center}
\end{figure*}

\begin{figure*}[ht]
\begin{center}
\includegraphics[width=1\textwidth]{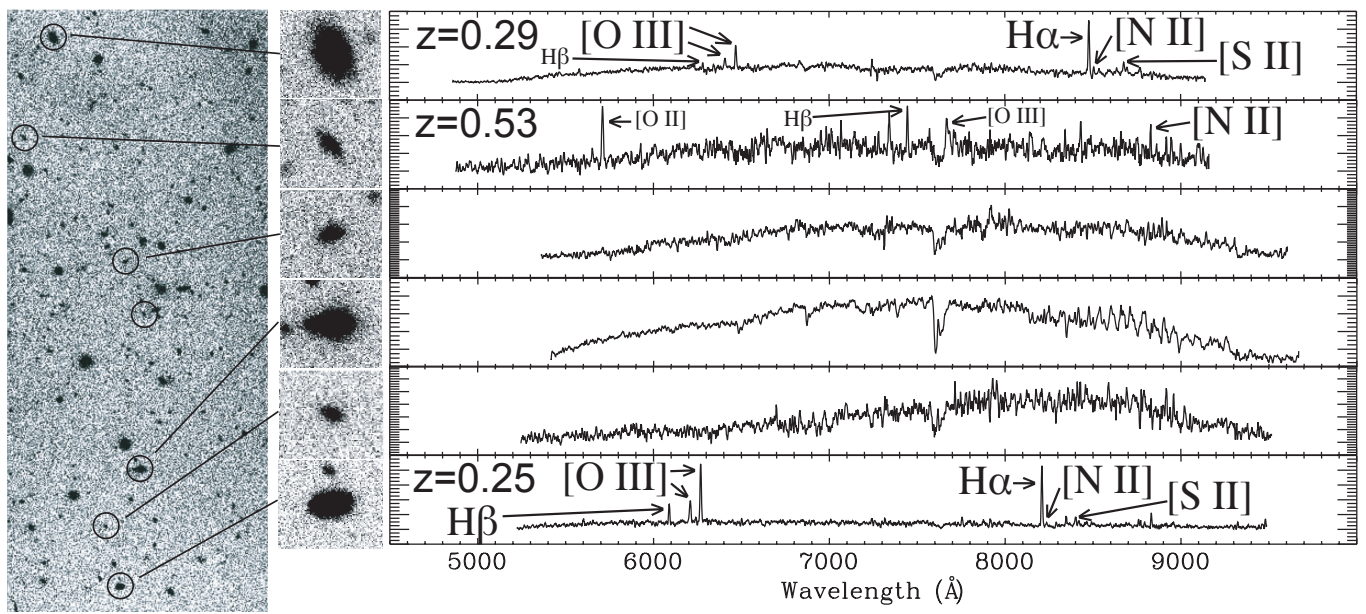}
\caption{The Gemini GMOS Nod \& Shuffle multi-object spectra aside from the host of LGRB 051022.  (The LGRB 051022 host location is indicated on the large image as the circle without a line connecting with a spectrum).  An eight slit mask was employed in MOS configuration, one slit being the object, another slit failed to yield a detectable spectrum and the other six lack any features of a redshift similar to the host of LGRB 051022. On three of the spectra the annotated lines were identified and place the objects at the labeled redshifts. \label{mos_spectra}}
\end{center}
\end{figure*}

Spectral extraction was performed with IRAF task ``apall" along with a matched extraction on an arc spectrum subsequently used for wavelength calibration.  The process yielded a spectrum with a spectral resolution of 1.37 {\AA} per pixel and a spatial resolution of 0.15 arc seconds per pixel.  The host galaxy spectrum contains several bright emission lines placing it at a redshift of $z=0.806$ consistent with the previous redshift measurements given by \cite{GCN4156} and \cite{CastroTirado}.  
The equivalent widths of several lines identified in the host galaxy spectrum are shown in Figure \ref{spectra} and listed in Table \ref{eqws}. None of the additional MOS objects showed features placing them at a similar enough redshift for a cluster association. The additional multi-object spectra collected are shown in Figure \ref{mos_spectra}.

\subsubsection{Host Near Infrared Spectroscopy}

Near infrared spectroscopic observations of the LGRB 051022 host were obtained with \textit{NIRSPEC} on the Keck II telescope on October 23$^{th}$ 2007.  
Our observations consisted of four 900 second exposures in the NIRSPEC-2 filter, using a 0.76 arc-second slit, and giving a spectral coverage from 1.09 to 1.29 $\mu$m.

\begin{figure}[h]
\begin{center}
\includegraphics[width=.48\textwidth,height=.3\textwidth]{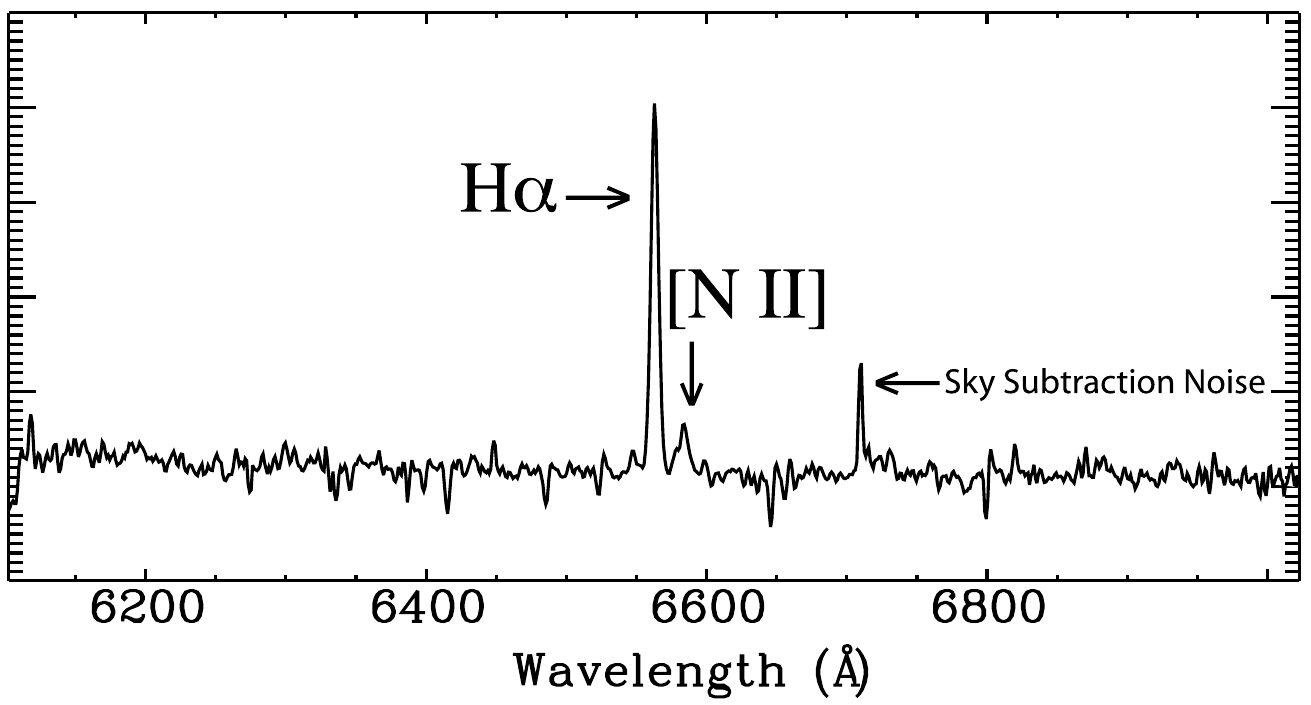}
\caption{Keck NIRSPEC near infrared spectrum of GRB 051022 shifted to the rest frame.  The [N II] to H$\alpha$ ratio is used to break the degeneracy of the R$_{23}$ method. \label{nirspectra}}
\end{center}
\end{figure}

Individual \textit{NIRSPEC} exposures were reduced using the standard procedure described in the online documentation from the NIRSPEC manual.\footnote{\protect{\url{http://www2.keck.hawaii.edu/inst/nirspec/redux.html}}}  The  object's placement on the slit was dithered between two locations with two exposures each, so that the combined image from each placement could be subtracted from the other to remove sky features.  
Due to the subtraction between the two dither placements, this image (like those described in the Nod \& Shuffle process perviously) contains two spectra, one of which is inverted.  Similarly, an inverted copy of the image is created, shifted to align the now positive spectrum with the positive spectrum on the original and then coadded, to yield a single combined spectrum.

Spectral extraction was performed with IRAF task ``apall" along with another extraction on non dither subtracted data to generate a sky spectrum that was used for wavelength calibration.  The process yielded a spectrum with a spectral resolution of 1.92 {\AA} per pixel and a spatial resolution of 0.16 arc seconds per pixel. The resulting spectrum with identified H$\alpha$ and 6583 {\AA} [N II] lines is shown in Figure \ref{nirspectra}.

The continuum was fit with a high order polynomial and the H$\alpha$ line with a Gaussian (with the continuum parameter set to the value returned from the polynomial fitting).  Similar Gaussian fitting on the 6583 {\AA} [N II] line was suboptimal and returned an inconsistent line width with a high fractional error.  In order to to obtain a more accurate result, the width of the H$\alpha$ line (with a much lower fractional error) was assumed for the 6583 {\AA} [N II] line and the line fit only for height (fitting also for center did not produce a statically different result in flux).  The fitting results are shown in Table \ref{eqws}.  Due do the proximity of the two lines the continuum is effectively unchanged between them thus, a direct flux ratio of the two lines can be calculated with lower error than would be naively expected from propagating the errors for the equivalent widths though the ratio and yields a value of 6.57 $\pm$ 0.52 for the H$\alpha$ to 6583 {\AA} [N II] line flux ratio.


\begin{table}[ht]
\begin{center}
\caption{\label{eqws}}
\vspace{-0.1 cm}
\begin{tabular}{ccccccccccccc}
\hline
\hline
        Line    &   Rest Wavelength (\AA) & Equivalent width\\
\hline

[O II] &   3727  & -67.00 $\pm$ 4.80\\

[Ne III] &   3869  & -8.00 $\pm$ 1.56\\

H$\gamma$ &   4340  & -10.00 $\pm$ 3.34\\

He II &   4686  & -4.86 $\pm$ 0.98\\

H$\beta$ &   4861  & -25.29 $\pm$ 4.85\\

[O III] &   4959  & -22.24 $\pm$ 4.55\\

[O III] &   5007  & -59.57 $\pm$ 6.36\\

H$\alpha$ &   6563  & -104.99 $\pm$ 4.09\\

[N II] &   6583  & -15.97 $\pm$ 1.37\\

\hline
\end{tabular}
\end{center}
\vspace{-0.2 cm}
Spectral line rest frame equivalent widths for the LGRB 051022 host.
\end{table}

\subsection{GRB 020819B}

\subsubsection{Burst Detection and Localization} \label{020819B_localization}

LGRB 020819B was initially detected by all three instruments on the HETE-2 satellite as well as the Gamma Ray Burst instrument in the Ulysses spacecraft.  Ulysses observations indicated a moderately long burst duration of approximately 20 seconds securely identifying this as a long gamma ray burst (\citealt{GCN1507} GCN 1507).  \cite{GCN1842} (GCN 1842) discovered a fading radio afterglow in VLA observations of the region.  \cite{GCN1844} (GCN 1844) found a clearly resolved galaxy with R $\sim$ 19.8 mag coincident with the radio afterglow position.   \cite{Jakobsson} measured a redshift of z=0.41 for the GRB 020319B host and observed that the radio afterglow is superposed on a faint structure they term ``the blob" located around 3 arcsecs from the galaxy center.  No optical or NIR afterglow was detected to a limiting magnitude of R $>$ 22 (\citealt{GCN1844} GCN 1844) and K' $>$ 19.5 mag \citep{Klose}, thus classifying LGRB 020819B as a ``dark" burst.  See Table \ref{detection020819B} for details of prompt and afterglow localizations.

\begin{table*}[ht]
\begin{center}
\caption{\label{detection020819B}}
\vspace{-0.1 cm}
\resizebox{\textwidth}{!} {
\begin{tabular}{ccccccccccc}
\hline
\hline
        $\Delta$T (days) & RA    &   Dec  & Error & Instrument & Band & Reference \\
\hline
0      & 23$^{h}$27$^{m}$07$^{s}$ & +6$^{\circ}$21$^{m}$50$^{s}$     & 7'  & HETE-2 WXM & 2-25 keV & \citealt{GCN1508} GCN 1508\\
0      & 23$^{h}$27$^{m}$24$^{s}$ & +6$^{\circ}$16$^{m}$08$^{s}$     & 2.5' & HETE-2 SXC & 0.5-10 keV & \citealt{GCN1508} GCN 1508\\
0      & 23$^{h}$27$^{m}$19.5$^{s}$ & +6$^{\circ}$17$^{m}$05$^{s}$     & 82" & HETE-2 SXC refined & 0.5-10 keV & \cite{villasenor}\\
1.75 & 23$^{h}$27$^{m}$19.475$^{s}$ & +6$^{\circ}$15$^{m}$55.95$^{s}$ & 1" & VLA & 8.46 GHz & \cite{GCN1842} GCN 1842\\
\hline
\end{tabular}}
\end{center}
\vspace{-0.2 cm}
Prompt and afterglow localizations of LGRB 020819B.  The SXC burst position was subsequently refined in \cite{villasenor}.  This burst was also detected by the Gamma Ray Burst instrument in the Ulysses spacecraft.  Using differential time of arrival from the two spacecraft the InterPlanetary Network (IPN) located the burst to a 4' wide annulus centered at RA: 345.048$^{\circ}$ Dec: -43.013$^{\circ}$ radius: 49.664$^{\circ}$ (\citealt{GCN1507} GCN 1507).
\end{table*}

\begin{figure}[h]
\begin{center}
\includegraphics[width=.48\textwidth]{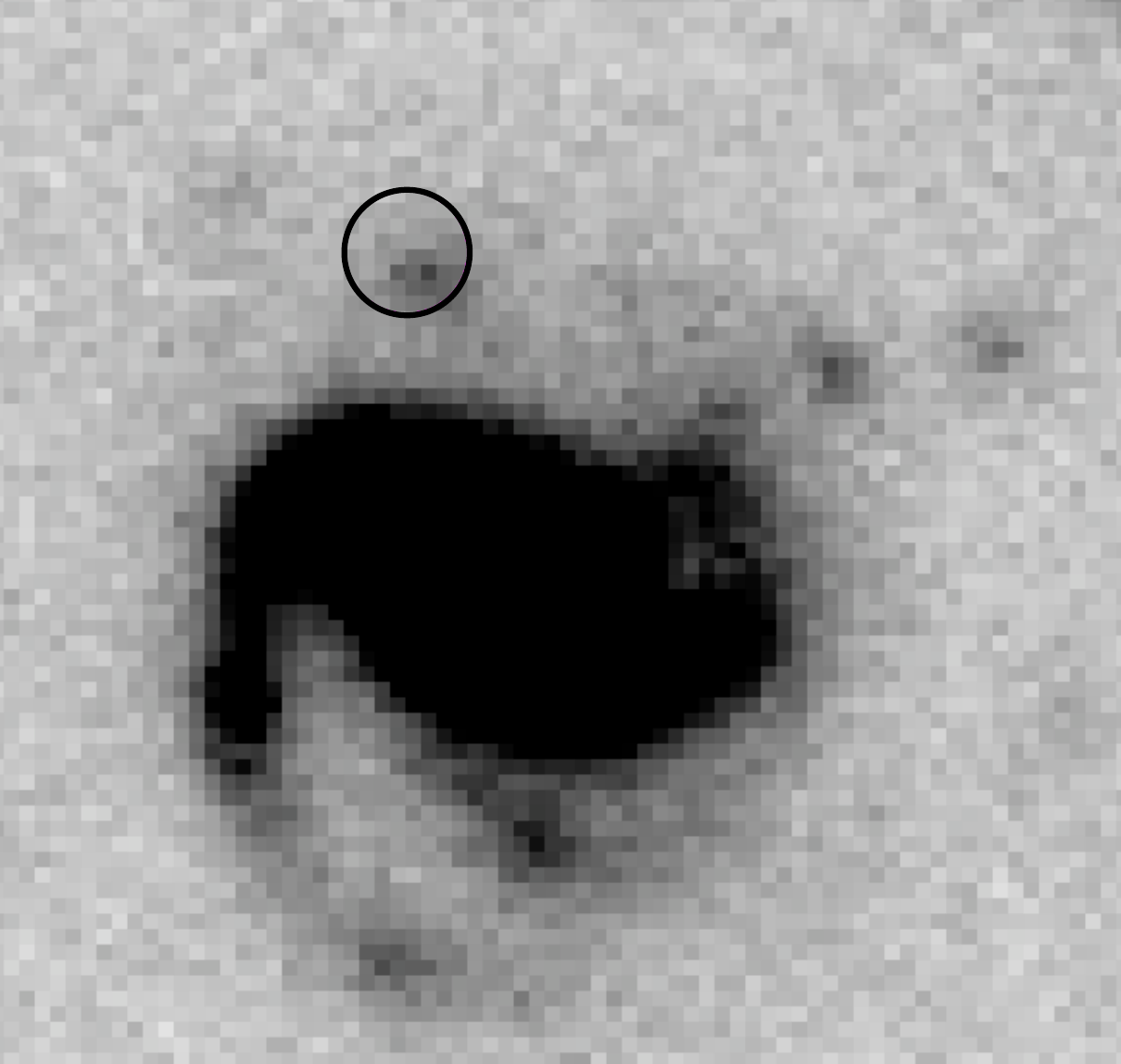}
\caption{Gemini North GMOS image of the host of GRB 020819.  The 1" diameter radio afterglow error circle is shown.  Unfortunately the resolution of the image and the diameter of ``the blob" are both about 0.6 arc-seconds.  While our spectroscopy places ``the blob" and the rest of the host at the same redshift, greater resolution than that obtainable from the ground is needed to determine whether ``the blob" is part of the spiral, an accreting galaxy like the Sagittarius Dwarf, or a satellite galaxy like the LMC.\label{020819B_image}}
\end{center}
\end{figure}


\subsubsection{Burst Host Imaging \label{020819B_phot}}

We acquired three dithered 600 second Gemini GMOS \textit{r} band exposures of the LGRB 020819B host on August 30th 2008 under photometric conditions to obtain high resolution imaging (program GN-2008B-Q-99). High resolution imaging is necessary for an accurate slit placement and it also provides a better understanding of the dynamics of the system. The photometric data was reduced with the standard Gemini GMOS IRAF\footnote{IRAF is distributed by the National Optical Astronomy Observatories, which is operated by the Association of Universities for Research in Astronomy, Inc., under cooperative agreement with the National Science Foundation.} packages and combined with a simple cosmic ray rejecting sum.  The Gemini GMOS image of the neighborhood of the radio afterglow of GRB 020819B along with the faint ``blob" is shown in Figure \ref{020819B_image}.  Aperture photometry yields an \textit{r} band magnitude of 24.20 $\pm$ 0.15 for ``the blob" and 19.829 $\pm$ 0.010 for the entire host galaxy system using the Gemini GMOS photometric zero points.  The host magnitudes agree well with the SDSS galaxy magnitude of 19.83 $\pm$ 0.04 and is reasonably consistent with the \citealt{Jakobsson} \textit{R} band value of 19.5 mag after uncorrecting for the galactic extinction of E(B-V) = 0.070 and converting to SDSS magnitude system.


For GRB 020819B, we estimate an absolute \textit{B} band magnitude of  -21.92 $\pm$ 0.01 for the host galaxy and -17.55 for ``the blob" using a similar approach as discussed in section \ref{051022host}.
The galaxy is unusually luminous for an LGRB host and belongs at about 3  $\times$ L* on the Schechter luminosity function.  "The blob" has an absolute magnitude approximately equal to the Large Magellanic Cloud (LMC -17.86  $\pm$ 0.05 using \cite{LMCmag} and \cite{LMCdist} for the LMC magnitude and distance values respectively) and about half the LMC's linear size.  


Our imaging resolution and the diameter of ``the blob", both being about 0.6 arc-seconds, makes it almost impossible to determine whether ``the blob" is an outlying part of the spiral, an accreting galaxy like the Sagittarius Dwarf, or a satellite galaxy like the LMC.  The asymmetry apparent in the disk of the spiral galaxy suggests that the systems are indeed interacting.
While HST imaging is required to definitively resolve the morphological nature of the LGRB progenitor region, a merging system was the operating assumption in selecting this burst as an analogous system to that of LGRB 051022 and ``the blob's" high star-formation rate supports (but does not prove) this.

\subsubsection{Burst Host Spectroscopy}

Spectroscopic observations of the spiral host galaxy of GRB 020819B were obtained at two separate epochs. The first epoch observations of the host galaxy nucleus were obtained with the Low Resolution Imaging Spectrograph (LRIS; \citealt{LRIS}) on the Keck I telescope on November 2nd, 2008. We acquired a second epoch spectrum of the star-forming region associated with the radio afterglow of GRB 020819B on November 19th, 2009. The results of the spectroscopic observations are presented in \cite{Levesque020819B}.  As noted in \cite{Levesque020819B} technical problems prevented the detection of 4861 {\AA} H$\beta$ and 4959 \& 5007 {\AA} [O III] lines on ``the blob" spectrum.




\section{Analysis of Host Galaxy Metallicities}

\subsection{Methodology}

\label{metal_measurement}

Since oxygen is the most abundant contributor to metallicity and metallicity is dominated by the combined abundance of oxygen, carbon and nitrogen the latter two of which share an approximately similar formation history to oxygen, oxygen is an ideal proxy of the overall metal abundance.  Oxygen is also typically the brightest contributor of metal lines in a spectrum and has no shortage of lines in the optical range making it a useful tracer.  Emission line diagnostics are specifically measuring the metallicity of H II regions, when applied to other galaxies this is used extensively as a proxy for the overall metallicity.  In specific application to the metallicity of bright stars, such as LGRB and core collapse supernovae progenitors, this is even more apt since a high degree of ionizing radiation (from the progenitor and similar neighboring stars) is virtually unavoidable and thus it is to be expected that the surrounding environment is an H II region.  This is doubly true in the former case as LGRB's occur in the brightest, most star-forming regions of their hosts \citep{Fruchter}.  Thus, any potential bias between the host galaxy's overall metallicity and the overall metallicity of its H II regions works in favor of a more accurate estimate of the LGRB progenitor's metallicity.  However, as galaxies are not uniform objects, the remaining issue is then how well the overall metallicity of the host galaxy's H II regions serves as a proxy for the specific metallicity of the LGRB progenitor's H II region (and thus the metallicity of the progenitor itself).


\subsubsection{$R_{23}$ Diagnostic\label{R23}}

The R$_{23}$ method is a commonly used metallicity diagnostic based on the electron temperature sensitivity of the oxygen spectral lines, achieved via using the ratio of Oxygen line strength to a spectral feature independent of metallicity.  In particular [O III] has lines at 4959 and 5007 {\AA} and [O II] has a particularly strong, typically unresolved, doublet at 3727 {\AA}. The metallicity independent 4861 {\AA} H$\beta$ line's convenient placement between the aforementioned Oxygen lines, gives the equation for R$_{23}$ used in the classical application of this diagnostic, see equation \ref{eqn}.  First proposed by Bernard Pagel in 1979 \citep{Pagel1979, Pagel1980}, R$_{23}$ has become the primary metallicity diagnostic for galaxies at $z>0.3$ (especially those where the faint [O III] 4363 {\AA} line is not measurable).

\begin{equation}
\label{eqn}
 R_{23} = {I_{3727} + I_{4959} + I_{5007} \over I_{H\beta}}
\end{equation}




In order to correlate the flux of a line belonging to an individual atomic ionization level with the total abundance of that element it is necessary to know what fraction of the element is ionized to the level in question.  This is achieved by measuring the flux ratio between the [O II] and [O III] line strength, giving the relative population in the O II and O III ionization states, and fitting the metallicity for that specific ionization state ratio.

Thus in the classical application of this diagnostic the R$_{23}$ value would be calculated from the measured line ratios and then compared along an [O III] to [O II] line ratio contour against the best calibration data available. This classical application however treats ionization as a parameter independent of metallicity and ignores the feedback the latter has on the former.  \cite{kd2002} solved this issue by using iterative fitting to dynamically factor the effects of the metallicity on the ionization parameter (without calculating the R$_{23}$ value shown above).  \cite{KobulnickyPhillips} established that the R$_{23}$ method can be directly used on equivalent width values (instead of reddening corrected flux values) and is found to be more accurate than flux ratios when reddening information is not available.

The R$_{23}$ method diagnostics suffer a degeneracy due to different effects being dominant at different regimes.  In the low metallicity regime the effects of the metals on the electron temperature of the system can be ignored due to their low relative abundance.  Thus the more metals in the electron gas, the more collisional excitations and more resultant flux in the metal lines.  As the metallicity rises however, emission from infrared fine-structure lines becomes significant and serves as a cooling mechanism, lowering the electron temperature, the electron velocity, the number of collisional excitations, and the metal line flux.

This temperature dependence causes two metallicity values (one high, one low) to generate the same metal line flux.  Unless one of the values can be obviously excluded or the two values are within the error range of the R$_{23}$ calibration, new empirical data is the only accurate way to break the degeneracy.  This is usually facilitated, as we have done here, by application of the [N II]/H$\alpha$ diagnostic described in section \ref{NII_Halpha}.  This has the obvious disadvantage of requiring additional measurements considerably outside the R$_{23}$ wavelength range often requiring a separate observation and for higher redshift objects a separate, near-infrared, instrument.

\subsubsection{[N II]/[O II] Diagnostic \label{NII_OII}}

The [N II]/[O II] ratio provides a diagnostic characterized in \cite{kd2002}.  Notably this diagnostic is relatively insensitive to differences in the ionization parameter and lacks the degeneracy issues plaguing R$_{23}$.  Given the wide separation between the 3727 {\AA} [O II] and 6584 {\AA} [N II] lines accurate flux calibration and measurement of the reddening are obvious issues and tend to limit this application.  Also the diagnostic's low variability to metallicity below half solar constrains its use to abundances above that, however as the [N II] line flux decreases with metallicity, for all but the most well measured spectra the low signal to noise ratio on the [N II] line flux will likely be the limiting factor.

Unlike R$_{23}$, [N II]/[O II] has not been tested using equivalent width values in place of fluxes.  However as the spectral energy distribution of the bluest galaxies almost never exceeds that of a flat continuum (in erg cm$^{-2}$ s$^{-1}$ {\AA}$^{-1}$), and extinction will preferentially suppress the 3727 {\AA} [O II] flux over the 6584 {\AA} [N II] flux, we can assume that the equivalent width ratio will function as an upper limit on the metallicity.  While we experiment with using this upper limit here, with encouraging results, our usage should be considered only anecdotal and a detailed study with statistical rigor and full characterization of errors is needed before its adoption is considered as a hard limit.

\subsubsection{[N II]/H$\alpha$ Diagnostic \label{NII_Halpha}}

The [N II]/H$\alpha$ ratio (also characterized in \citealt{kd2002}) provides a crude metallicity indicator however due to its strong dependence on the ionization parameter it provides only a gross estimate of abundance unless the ionization in known.  Since measure of the ionization parameter requires measurement of lines which themselves constitute better metallicity diagnostics application of ionization parameter correction to this diagnostic is of limited utility.  Also the diagnostic is easily distorted by contamination from shock excitation and AGN rendering it comparatively inaccurate.  Thus its primary application is selecting between the degenerate upper and lower branch values provided by the R$_{23}$ diagnostic (as stated in section \ref{R23}).  



\subsubsection{Scale \& Code \label{code}}

Due to differences between metallicity diagnostics and their various calibrations, a true comparison of metallicity requires using a common scale and if possible a consistent methodology and diagnostic.  Here we adopt the \cite{KobulnickyKewley} scale.  Based on the \cite{solar} 6300 {\AA} [O I] line measurements of the Sun, solar metallicity is estimated to be log(O/H)+12 = 8.69 $\pm$ 0.05 in this scale.  It should be noted that while the \cite{kd2002} diagnostics listed above can be cross calibrated to a high degree of accuracy due to the large number of H II regions where all the diagnostics can be applied this is not true of the density-sensitive 6300 {\AA} [O I] line measurement where the line strength is insufficient for widespread application.  Thus the \cite{KobulnickyKewley} scale is more accurate internally than to any absolute reference to solar value and such comparisons should be limited to broad generalizations.  For conversion to other scales and discussion of associated issues we refer the reader to the transforms in \cite{KewleyEllison}.

To ensure consistency of method, we employ an improved version of the idl code outlined in \cite{kd2002} (updated to the \citealt{KobulnickyKewley} scale) to calculate metallicity values, including recalculation of comparison values from published fluxes.  Since the Kewley code applies ionization across all diagnostics and was used to cross calibrate the various diagnostics with each other, we expect this methodology to give the best cross diagnostic agreement.  Still, when possible and unless otherwise specifically described, the R$_{23}$ diagnostic output is the metallicity value adopted.

\subsection{LGRB 051022 \label{051022metal}}

Using the [O II], [O III], and H$\beta$ host galaxy line equivalent widths from our Gemini GMOS South spectroscopy yields degenerate R$_{23}$ metallicity values of  log(O/H)+12 = 8.18 and  log(O/H)+12 = 8.77 (about 1/4 solar and slightly super-solar respectively).
The 6583 {\AA} [N II]  and 6563 {\AA} H$\alpha$ line values (obtained with NIRSPEC on Keck II due to their being red-shifted into the infrared)  constrains the R$_{23}$ degeneracy to the upper branch yielding a host galaxy metallicity of log(O/H)+12 = 8.77 $\pm$ 0.07 in the \cite{KobulnickyKewley} scale
 which was the highest measured metallicity of any long burst host galaxy then seen. We adopt this metallicity value for the system subsequently in the paper.

Interestingly however application of the [N II]/H$\alpha$ diagnostic yields a value of log(O/H)+12 = 8.90 notably higher than with R$_{23}$.  Application of the [N II]/[O II] diagnostic with equivalent widths assuming the flat continuum blue constraint outlined section \ref{NII_OII} gives a soft upper limit of log(O/H)+12 = 8.84 on the metallicity.  The [N II]/H$\alpha$ diagnostic is known to be comparatively inaccurate \citep{kd2002}, even when it is corrected with the ionization parameter determined from the [O II] to [O III] line ratio. Still the higher  [N II]/[O II] diagnostic value may be indicative of either a  nitrogen over abundance in the system (respective to oxygen), ionization parameter differences between the slit placement in the GMOS and NIRSPEC observations, or a metallicity variation across the merging host system.

Additionally \cite{Levesque051022} obtained independent optical spectroscopy of the GRB 051022 host with Keck LRIS and determined an R$_{23}$ metallicity of log(O/H)+12 = 8.62 using flux values.  \cite{Levesque051022} use the H$\beta$ to H$\gamma$ ratio to determine a reddening of E(B-V) = 0.50 and our [N II]/H$\alpha$ ratio to resolve degeneracy.  Based on their published flux values we recalculate an R$_{23}$ value of log(O/H)+12 = 8.64 (via the methodology in section \ref{code}).  Applying that ionization parameter to the [N II]/H$\alpha$ ratio \& diagnostic yields a value of log(O/H)+12 = 8.83 again notably higher than with R$_{23}$.  We adopt the reddening value from \cite{Levesque051022} to attempt a crude flux calibration of the 6583 {\AA} [N II] line via setting the H$\alpha$ flux at its expected value with respect to the H$\beta$ line from the Balmer decrement.  Following this by applying the observed  [N II]/H$\alpha$ ratio gives an [N II]/[O II] value of log(O/H)+12 = 8.67 for the system.  The values or limits available for the various diagnostics in GRB 051022 host system using both fluxes and equivalent widths are listed in Table \ref{metallicities}. For LGRB 051022 the fluxes are the published values from \cite{Levesque051022} which were obtained via separate optical spectroscopy and suspected to be of a separate region of the host.

While a difference between the two independent R$_{23}$ observations of only 0.13 dex might seem minor it is almost twice the intrinsic $\pm$0.07 dex error of the \cite{KobulnickyKewley} diagnostic. The surprisingly good agreement between the R$_{23}$ and [N II]/[O II] diagnostics applied to the \cite{Levesque051022} LRIS spectrum suggests that our Keck NIRSPEC spectroscopy might more closely match the LRIS spectrum than our GMOS spectroscopy and highlights that the difference may lie in the position angles used.  Our multi-object spectroscopy of the host was constrained to a slit position angle of 90$^{\circ}$ so as to match our GMOS imaging from which the slit mask was derived (which itself was oriented to a position angle to obtain a bright On Instrument Wave Front Sensor [OIWFS] star while still orienting the image with respect to a cardinal direction).  This places the slit axis almost exactly perpendicular to the axis between the northern and southern regions in Figure \ref{hst_initial} and maximizes the likelihood of preferentially obtaining one of the regions over the other.  Similarly, the Keck LRIS spectroscopy of \cite{Levesque051022} used a slit a position angle of 105$^{\circ}$ (private communication) also placing their slit axis roughly perpendicular to the extended direction of the host.  Our Keck NIRSPEC observations however used slit a position angle of 210$^{\circ}$ to orient the slit along the apparent length of the galaxy (in the ground based imaging).  While the separation between the two regions is only about 0.6" and thus a large amount of contamination is to be expected even if the slit were perfectly covering only one of the regions this only implies that any apparent metallicity differences seen between the two spectra are likely a significant underestimate of those in the actual system.

\subsection{LGRB 020819B}

\subsubsection{Host Galaxy Nucleus}

\cite{Levesque020819B} give metallicity values of log(O/H)+12 = 9.0 and 8.8 for the [N II]/[O II] and [N II]/H$\alpha$ diagnostics respectively.  For maximum possible consistency with other objects we recalculate these values using the code and procedure outlined in section \ref{metal_measurement}.  This yields values of log(O/H)+12 = 8.97, 8.98, and 8.99 for the R$_{23}$, [N II]/[O II] and [N II]/H$\alpha$ metallicity diagnostics respectively.  (The primary difference in methodology being the correction here of galactic extinction based on the \citealt{Schlegel} dust maps before correcting for internal extinction via the Balmer decrement.  In this as most cases this difference is quite minor but is employed for consistency).  The R$_{23}$ value is adopted as the metallicity of the LGRB 020819B host galaxy.


\subsubsection{``The Blob" outlying region hosting the LGRB} \label{blob_metal}

Given the localization of the burst to an outlying structure the properties of the host galaxy nucleus are likely irrelevant to understand the GRB progenitor environment (but useful in understanding sources of potential error on host studies where such positioning is not available or where the system geometry does not allow for such a clean identification of a progenitor region, i.e. an edge on system).  Thus direct metallicity measurement of ``the blob", outlying the galaxy where the burst occurred, is necessary for inferring the GRB progenitor properties.

This process is greatly complicated by the failure to obtain the H$\beta$ and [O III] lines \citep{Levesque020819B}.  The preferred R$_{23}$ metallicity diagnostic obviously cannot be applied at all.  The [N II]/[O II] diagnostic while itself unaffected is essentially useless without knowledge of the extinction (obtained by fitting the observed Balmer lines to expected ratios) required to correct the observed flux values for reddening.  The [N II]/H$\alpha$ diagnostic is virtually unaffected by reddening (due to the close proximity of the two lines) however it is highly dependent on the ionization parameter (typically determined from the [O III] to [O II] line ratio).  Still the [N II]/H$\alpha$ diagnostic is the best estimate possible from the flux values available and yields a metallicity of log(O/H)+12 = 8.95 for ``the blob."  This value and those for the host galaxy nucleus are consistent with the metallicity values given in \cite{Levesque020819B} shifted to the \cite{KobulnickyKewley} scale using the transformations in \cite{KewleyEllison}.  However without knowledge of the ionization parameter the [N II]/H$\alpha$ diagnostic at best a crude estimate of the metallicity.

In order to quantize constraints on [N II]/H$\alpha$ diagnostic error we adopt a more involved approach.  Given that the H$\alpha$ flux is known and the intrinsic H$\alpha$ to H$\beta$ ratio set by the Balmer decrement, the observed H$\beta$ flux is a function of the extinction.  Similarly as the 3727 {\AA} [O II] flux is known and the [O II] to [O III] line ratios are a function of the ionization, the flux of the [O III] lines is a function of the ionization parameter.  Thus the problem can be decomposed into three metallicity diagnostics ([N II]/[O II], R$_{23}$, and [N II]/H$\alpha$) and three unknown parameters (extinction, [O III] flux, and metallicity).  The H$\beta$ flux is estimated from the H$\alpha$ flux and extinction using the Balmer decrement.  The system can thus be modeled by iterating though the reasonable extinction and [O III] flux range and seeking concordance between the three metallicity diagnostics. The iteration yields three solutions of E(B -- V), 5007 {\AA} [O III] flux and metallicity as listed in Table \ref{iterated}.  

\begin{table*}[ht]
\begin{center}
\caption{\label{iterated}}
\vspace{-0.1 cm}
\begin{tabular}{cccccccccc}
\hline
\hline
        Extinction    &   Extinction derived & Estimated  5007 {\AA} & Metallicity\\
        E(B -- V)    &   H$\beta$ flux   & [O III] flux  &  log(O/H)+12\\
        (magnitudes)    & 10$^{-17}$ erg cm$^{-2}$ s$^{-1}$ & 10$^{-17}$ erg cm$^{-2}$ s$^{-1}$  & \cite{KobulnickyKewley} scale\\
\hline

0.31 &  9.3 & 4.7 &  9.08\\

0.67 & 6.2 & 4.3 & 8.93\\

1.09 &  3.9  & 1.5 & 8.70\\

\hline
\end{tabular}
\end{center}
\vspace{-0.2 cm}
Iterated multi-metallicity diagnostic solutions. The flux values are given as they would have been measured (i.e. prior to correcting for extinction).
\end{table*}

\begin{figure}[h!]
\begin{center}

\includegraphics[width=.48\textwidth]{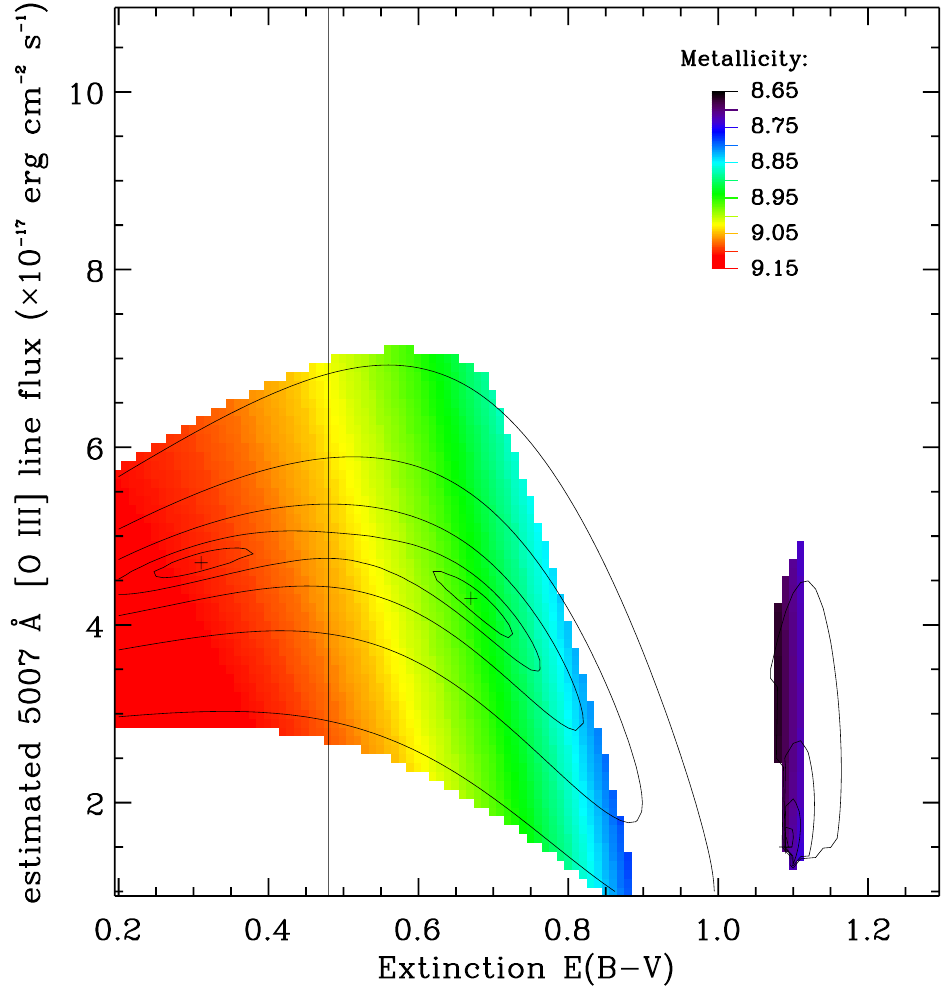}
\vspace{-10pt}
\caption{\label{metal_fit}  Allowed multi-metallicity diagnostic solutions for given extinction and [O III] flux iterated values for ``the blob" region of the LGRB 020819B host.  Fill color shows the R$_{23}$ metallicity of the given solution and the contours show the standard deviation between the different metallicity diagnostics.  The three metallicity diagnosis ($R_{23}$, [N II]/[O II], \& [N II]/H$\alpha$) and the three unknowns (i.e. the two missing lines 4861 {\AA} H$\beta$ \& 5007  {\AA} [O III] and the metallicity) effectively form a three equations and three variables problem.  Thus we iterate across reasonable extinction and 5007 {\AA} line flus values discarding those solutions where the $R_{23}$ and [N II]/[O II] diagnosis differ by more than 0.05 dex or the mean of those differ from the [N II]/H$\alpha$ by more than 0.15 dex (approximately the optimal cross diagnostic agreement observed).  Though not plotted we iterated out to an extinction of E(B -- V) of 2 without any additional allowed values.  Standard deviation contours of 0.005, 0.01, 0.02, 0.04, \& 0.08 are plotted with a + marking the point of exact agreement (i.e. where all three metallicity diagnostics give the same value).  Note that the allowed solution (the area in the figure with a shown metallicity color) cutoff does not follow any of these contours due to its being determined exactly as described with different criteria on agreement between the different diagnostics.  Additionally, a vertical line shows the extinction lower limit value of E(B -- V) = 0.48 discussed in section \ref{blob_metal}.  Given the wide range of allowed metallicities even with optimal cross diagnostic agreement we conclude that the metallicity of the LGRB 020819B hosts ``the blob" region cannot be constrained much beyond the upper branch of the R$_{23}$ diagnostic.  Thus while the specific region termed ``the blob"  hosting LGRB 020819B is at much higher metallicity than typical LGRB hosts, comparison of its metallicity to the center of its host galaxy cannot be determined from the data yet obtained.}

\end{center}
\end{figure}

The three cases in Table \ref{iterated} represent the zero error cases only.  In reality disagreement between the metallicity values given by the different diagnostics must be considered.  In figure \ref{metal_fit} the diagnostics disagreement is shown with the overplotted contours or representing the degree of difference between the different metallicity diagnostics (the + marks the three cases where the diagnostics agree exactly given in Table \ref{iterated}).  Assuming a conservative $\sim$0.1 dex error on the values  gives an approximate metallicity range for the [N II]/H$\alpha$ ratio and diagnostic at hand of about log(O/H)+12 = 8.67 to 9.18 or a value of log(O/H)+12 = 8.95$^{+0.22}_{-0.28}$ on the [N II]/H$\alpha$ diagnostic.


While the lack of an extinction value for ``the blob" makes it impractical to apply the [N II]/[O II] diagnostic to determine metallicity it can however be used to set some limits on the system.   The metallicity value decreases as one increases the extinction estimate. Thus for the zero extinction case ``the blob" metallicity of log(O/H)+12 = 9.18 is a hard upper limit.  Similarly adopting the flat continuum blue constraint (outlined in section \ref{NII_OII}) on the equivalent width values with the [N II]/[O II] diagnostic gives a soft upper limit of log(O/H)+12 = 9.01 for ``the blob" region.  This added [N II]/[O II] upper limit constraint removes the upper metallicity zero error case in Table \ref{iterated} and reduces the metallicity range from that allowed under the [N II]/H$\alpha$ diagnostic.  Additionally, adopting the equivalent width soft upper limit value and applying the [N II]/[O II] diagnostic with fluxes yields, where its metallicity converges at, an E(B -- V) = 0.48 lower limit on the extinction.  
This corresponds to an extinction of A$_V$ = 1.3 magnitudes or greater which is the criteria for modest extinction in \cite{Jakobsson} model 1 and thus potentially explains the ``dark" nature of this burst. The values or limits available for the various diagnostics in the GRB 020819B host nucleus and ``the blob" region using both fluxes and equivalent widths are given in Table \ref{metallicities}.



\begin{table*}[t]
\begin{center}
\caption{ \label{metallicities}}
\vspace{-0.1 cm}
\begin{tabular}{lcccccccc}
\hline
\hline
        Input    &   R$_{23}$ & [N II]/[O II]  & [N II]/H$\alpha$\\
\hline

LGRB 020819B host ``blob" fluxes &  ... & $\leq$ 9.18 & 8.95 \\

LGRB 020819B host ``blob" equivalent widths & ...  & $\leq$ 9.01 & 9.02\\

LGRB 020819B host nucleus fluxes & 8.97 & 8.98 & 8.99\\

LGRB 020819B host nucleus equivalent widths & 9.08 & $\leq$ 9.10 & 8.97\\

\hline

LGRB 051022 host system equivalent widths &  8.77 & $\leq$ 8.84 & 8.90\\

LGRB 051022 host system fluxes & 8.64 & 8.67 & 8.83\\

\hline
\end{tabular}
\end{center}
\vspace{-0.2 cm}
Metallicity values for various diagnostics and inputs.  (The LGRB 051022 host system fluxes are from the \cite{Levesque051022} spectrum which is distinct from the spectrum used for the LGRB 051022 equivalent widths). 
\end{table*}

\subsection{LGRB 050826}


In addition to LGRBs 020819B and 051022, a third case, LGRB 050826, was identified as being at high metallicity by \cite{Levesque2}.  Although we did not observe this object, a discussion of LGRB 050826 is in order due to its being the remaining high metallicity LGRB in the \cite{stats_paper} sample.  This object, with a host at log(O/H)+12 = 8.84 (recalculated using the published fluxes via the methodology described in section \ref{metal_measurement} and consistent with \citealt{Levesque2} value), is unique both in having an optical counterpart (\citealt{GCN4749} GCN 4749), and in being the first high metallicity LGRB host found without some prior expectation of this result (remember than both LGRBs 020819B and 051022 were suspected to be at high metallicity due to their unusually bright host galaxies).  Due to its discovery in such an untargeted manner and otherwise being a burst of no particular importance to the community, it is as yet quite poorly studied.

\begin{figure}[h!]
\begin{center}
\includegraphics[width=0.48\textwidth]{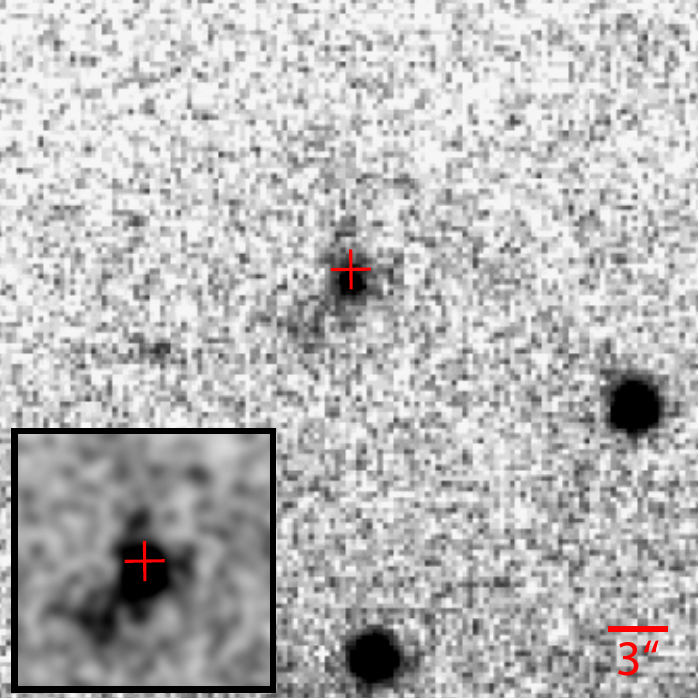}
\caption{\label{050826} Image of the LGRB 050826 host galaxy taken with RETROCAM on the MDM 2.4m telescope.  Inset is a heavily smoothed cutout of the same image at the same scale.  Note the irregular and likely dynamical host galaxy morphology.  Image from \texttt{http://user.astro.columbia.edu/$\sim$jules/grb/050826/}.}
\end{center}
\end{figure}

Unfortunately the host of LGRB 050826 is lacking any HST observations so it is difficult to study its properties in great detail --- see Figure \ref{050826} for an image of the host galaxy.  Morphologically it appears as the confluence of the properties of the other two high metallicity LGRB host galaxies and the larger (presumably predominately low metallicity) LGRB host sample of \citep{Fruchter}.  Like the other high metallicity LGRB hosts this galaxy is brighter than the typical host but consistent with the general luminosity metallicity relation for its redshift.  However the as with most LGRB hosts the galaxy is small (0.3 L*) and clearly irregular.  It is difficult to determine with the obtained image quality if the host is a merging system however this possibility certainly can not be excluded.  The burst itself is sub luminous consistent with the majority of LGRB events detected in the local population \citep{Mirabal}.  Not only does LGRB 050826 establish that a burst in a high metallicity environment can have an optical transient but also this is the first case where the metallicity was determined to be high without previous suspicion that this was likely the case before obtaining the metallicity measurement.

\section{High Metallicity LGRBs as a Population}\label{hzp}



The individual characteristics of the high metallicity LGRBs and their hosts warrant some consideration.  Given that these initial two high metallicity LGRBs (051022 \& 020819B) were both dark bursts, this led to early speculation of whether it was the dark nature of the burst that allowed it to occur in such a high metallicity environment  (i.e. dark bursts were a distinct phenomenon without the metal aversion seen in typical bursts with an optical afterglow) or whether the absence of an optical transient was simply a product of this burst's environment (i.e. larger degree of extinction due to geometry and greater prevalence of dust with metallicity).  The addition of LGRB 050826 with an optical transient strongly favors the latter argument and strongly suggests that further dark bursts will also be high metallicity.  Our observations of LGRBs 051022 \& 020819B are also consistent with the \cite{Perley2013} 
observation that dark burst host galaxies to be more massive, star-forming, dust obscured, and thus presumably metal rich than than LGRB hosts with optical afterglows.  This has obvious potential implications on the LGRB metal aversion debate.  However as current estimates are that no more than 20 to 30 \% of LGRBs are dark  \citep{Perley_dark_frac}, even assuming all dark LGRBs are high metallicity, there is still an insufficient fraction of such bursts to ascribe LGRB formation as a metallicity independent function of star-formation (even using the most generous estimates of as much as half the star-formation occurring in small metal-poor galaxies), thus there is still a strong selection effect toward low metallicity environments \citep{stats_paper, Perley2013}.


The high metallicity LGRB host morphology is noteworthy.  Our HST imaging conclusively shows that LGRB 051022's host is a merging system and our ground based Gemini imaging of LGRB 020819B is also shown a clearly disturbed morphology consistent with (but not conclusive of) a small galaxy (hosting the burst) being absorbed into the grand design spiral.  LGRB 050826 lacks deep resolved imaging and is described as possessing a "bright core and irregular extension" \citep{GCN4749}.  
While the association between high metallicity environments and bursts being dark could be as simple as the metallicity being necessary for the formation of sufficient dust to provide the required extinction the effect of morphology is more complicated.  Given that two of the three known high metallicity LGRBs are dark, and assuming that both cases are mergers, this suggests that this is likely more than a chance occurrence.  Combined with the fact that these two bursts have the highest star-formation rate of any LGRB hosts z $<$ 0.9 (approximately the redshift range of measured emission line host metallicity) suggests that extremely rapid self enrichment is occurring.

\begin{figure}[ht!]
\begin{center}

\includegraphics[width=.48\textwidth]{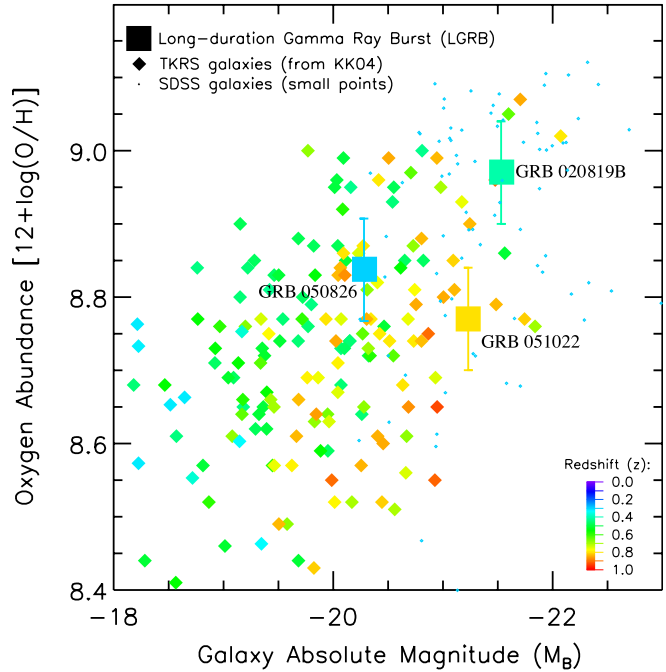}

\caption{\label{LZz_excerpt} Central galaxy metallicity versus\ B band absolute magnitude of high metallicity LGRB hosts (squares), TKRS galaxies (diamonds), and SDSS galaxies (small dots) with color used to index redshift.  For clarity, the SDSS galaxy population has been restricted to galaxies with a similar redshift to GRB 050826 (the other plotted LGRBs exceed the redshift range of the SDSS population).  Note that the three high metallicity LGRB hosts are consistent with the general galaxy population (SDSS \& TKRS galaxies combined) luminosity-metallicity trends for their redshift.  Thus the metallicities of the high metallicity LGRBs are fully consistent with typical star-forming galaxies of comparable brightness and redshift.}

\end{center}
\end{figure}

In \cite{stats_paper}, we analyzed the general metallicity trends of the LGRB population, compiling data from the two objects studied here as well as all other suiTable LGRB host spectra and other object classes for comparison and found that LGRBs have a strong intrinsic preference for low metallicity environments.  Of particular note is that the high metallicity LGRB population (of these three objects out of 14 total LGRBs) appears consistent with the general star-forming galaxy population of comparable brightness and redshift. 

\section{Discussion}


While the majority of the LGRB population is found in host galaxies far more metal-poor than typical galaxies of comparable luminosity and redshift, thus reflecting an intrinsic preference of LGRBs to occur preferentially in low metallicity environments, high metallicity LGRB hosts (while rare) do exist. Specifically, we identify the cases of LGRBs 020819B, 050826, \& 051022.  What is most notable about the three cases observed to date is that each of the LGRB host galaxies are consistent with the TKRS sample with no visible offset toward a lower metallicity (see figure \ref{LZz_excerpt}).  Given that there is some intrinsic distribution in galaxy metallicities, this result is  surprising --- where there an ongoing preference for low metallicity environments then even after selecting a set of LGRBs with high metallicity host galaxies one would still expect those LGRBs to have as low a metallicity environment as that selection allowed.  Even if an individual LGRB containing galaxy is at a high metallicity we would still expect it to be at a lower metallicity than typical for a galaxy with the same brightness and redshift.  Within a high metallicity galaxy we would expect to find the LGRB in a region of lower than typical metallicity.  However we find that all three of our high galaxies LGRB host galaxies to be at a brightness and redshift typical for the general galaxy population of \cite{stats_paper}.  Furthermore, we don't see any obvious signs that high metallicity LGRB are in the lowest metallicity parts of there hosts.  The location of LGRB 020819B looked prosing in this respect however metallicity measurement of the specific host galaxy region where it occurred excludes a typical low metallicity environment.  Without either sign of ongoing metal aversion in environments of higher metallicity than in which LGRBs are typically observed we therefore deduce that it is likely still possible to form an LGRB in a high metallicity environment although with greater rarity. \label{marginal}







In \cite{stats_paper}, we analyze LGRB population metallically distribution in detail and find that aside from our sample's 3 high metallicity hosts (out of a total population of 14) the remaining LGRB population is constrained to the lowest metallicity tenth of available star-formation.  This simple counting statistic is complicated by a number of factors.  First of all, 2 of the 3 high metallicity LGRB hosts were selected for metallicity measurement in a biased manner, deliberately searching based on prior information for cases that would likely violate the low metallicity LGRB host trend observed to date.  Secondly, the general galaxy mass metallicity relationship in concert with the greater observational practicality of measuring the metallicity of brighter galaxies and the greater range and thus sample size available therein combines to skew general LGRB host metallicity surveys (those without efforts to cherry pick anomalous cases) toward the higher metallicity end of the host population.  These selection biases are addressed in \cite{stats_paper} although we suspect the total effect of these biases is likely a factor of a few, at most.

Thus while the likelihood of forming an LGRB is decreasing with increasing metallicity, either asymptotically or linearly, the preference appears to be truncated at high metallicity such that the probability of forming an LGRB, instead of approaching zero approaches a small nearly (or perhaps actually) constant value. This LGRB metallicity dependance function leads to three potential interpretations.

\label{cases} (1) LGRBs do not occur in high metallicity environments and those seen in high metallicity hosts are in fact occurring in low metallicity environments that have become associated with otherwise high metallicity hosts but remain unenriched. The spatially resolved metallicity observations of LGRB 020918B seem to suggest that this is not the case. However due to the limits on the precision of the site metallicity measurement (due to detector problems), the limited spatial resolution of the dark burst, the lack of definitive morphology information on the burst progenitor region and the presence of only a single such spatially resolved high metallicity LGRB observation we find this evidence insufficient to be fully exclusionary.  Thus while this explanation cannot be completely ruled out, based on our work to date we believe it to be rather improbable.

(2) The LGRB formation mechanism while preferring low metallicity environments does not absolutely require it resulting in a continuous (though probably nonlinear) decline in burst formation with increasing metallicity with an observed sharp drop off around half solar metallicity.  The primary physical implication of this being that there is no physical limit on the metallicity allowed for LGRB formation only a degradation of the formation process with increasing metallicity.  Both the absence of a correlation between LGRB energetics and metallicity \citep{Levesque_No_Correlation} as well as the absence of any apparent bias in the high metallicity LGRB population toward lower than average metallicity (for galaxies of the same brightness and redshift) both suggest that this is not the case.  However the small number of such high metallicity LGRBs and the intrinsic noise in the metallicity measurements makes this evidence again insufficient to be exclusionary.

(3) The typical low metallicity LGRBs and the few high metallicity cases are the result of physically differentiable burst formation pathways with only the former affected by the metallicity of the burst environment and the later simply occurring much more infrequently.  Presumably such a metallicity insensitive pathway would also produce LGRBs at low metallicity (as an infrequent addition to those produced through more common metallicity sensitive route) in numbers predicable from extending the high metallicity LGRB rate per unit star-formation down to smaller metal-poor galaxies.  

The binary explosive common-envelope ejection mechanism of \cite{Podsiadlowski} provides a credible production scenario for short-period black-hole binaries with the bare carbon oxygen core progenitor likely needed for LGRB and type Ic supernovae production.  As noted therein this mechanism likely incorporates a low metallicity bias as the required red-supergiant branch (\citealt{Lauterborn1970} case C) mass transfer is more common at lower metallicity.  Whether this mechanism best explains only the high metallicity LGRBs (whose short period systems are not produced through typical pathways \citealt{Linden}) or all LGRBs is debatable however as only a small fraction of Ic supernovae become LGRBs we suspect in general that the introduction of a binary companion to LGRB formation, assuming it is not a prerequisite, would probably prove quite disruptive.  The metallicity bias in this formation mechanism would likely favor a continuing preference for low metallicity, even in the high metallicity range, which conflicts with the exceedingly tentative evidence (that at high metallicity marginal differences in galaxy metallicity do not have an appreciable effect) presented in section \ref{marginal} and favors the case 2 scenario in section \ref{cases}.

Overall, a binary formation model with a very particular and rare mass transfer process and bounds for LGRB production is well-suited to provide the necessary scarcity of LGRB events required to match the observed disparity between LGRB formation rates and those of non-LGRB broad-lined type Ic supernovae \citep{form_rate_letter}.  Considering the much less total coverage of present supernovae searches (compared with GRB detectors) the number of broad-lined type Ic supernovae found at their respective low redshifts likely significantly out paces LGRB losses due to non-detection of off axis bursts and the lower total star-formation in requisite low metallicity environments.  The additional requirement of a binary system close enough to allow mass transfer interaction of a specific rare type would thus introduce such an additional lowering of the LGRB formation rate.  Whether SNe progenitor rotation can provide the required one LGRB per $\sim$40 Type Ic-bl SNe under optimal low metallicity conditions \citep{form_rate_letter} degrading to one LGRB per $\sim$120 Type Ic-bl SNe at higher metallicities remains an open question \citep{Yoon2010}.

\section{Summary}

We present our imaging and spectroscopic observations of the host galaxies of two dark long bursts with anomalously high metallicities, LGRB 051022 and LGRB 020819B, which in conjunction with another LGRB event, LGRB 050826 (shown in figure \ref{050826}), with an optical afterglow \citep{Levesque2} comprise the three LGRBs with high metallicity host galaxies in the \cite{stats_paper} sample.   For both LGRB 051022 and LGRB 020819B no optical counterpart was detected however radio and X-ray source was detected classifying these as ``dark" LGRBs but allowing us to locate the bursts upon their host galaxies.  LGRB 051022's host is a z=0.8 extended object with two separated star-forming regions and a tidal tail (visible in figures \ref{hst_initial} and  \ref{tt_image} respectivly).  LGRB 020819B occurred in a small outlying structure termed ``the blob" on the outskirts of a z=0.4 grand design spiral galaxy (shown in figure \ref{020819B_image}).

We used the $R_{23}$, [N II]/[O II] and [N II]/H$\alpha$ diagnostics to calculate the metallicity in LGRB 051022 and LGRB 020819B.  The metallicity estimates for LGRB 051022 using different diagnostics are in reasonable agreement with each other, and slightly different than that given in \cite{Levesque051022} from independent optical spectroscopy however our discovery of a merging morphology for this host raises the issue of internal metallicity disparity and makes such ensemble metallicity measurements less than ideal.  For LGRB 020819B in addition to the host galaxy core we were also able to measure ``the blob" metallicity independently and do not find its metallicity to differ from the galaxy center within the error of our estimate though only with the considerably less sensitive [N II]/H$\alpha$ diagnostic  In an attempt to improve on this result (despite less than complete spectroscopy due to instrument failure) we use the lines we do have to create essentially a 3 equations (3 different metallicity diagnostics), with 3 variables problem (metallicity, extinction, and ionization), and naturally find three solutions (shown in figure \ref{metal_fit}).  One of which does allow for ``the blob" region to be at a significantly lower metallicity than the galaxy core but all solutions are still considerably higher in metallicity than that seen in typical GRB hosts.

In \cite{stats_paper}, we showed that LGRBs exhibit a strong and apparently intrinsic preference for low metallicity environments.  However, as we note therein, some exceptions do exist to this trend --- three of the 14 LGRBs in the sample possess abundances of about solar and above.  Not only do the three high metallicity LGRB hosts (051022, 020819B, and 050826) not share the typical low metallicities of LGRB hosts, they are consistent with the general star-forming galaxy population of comparable brightness \& redshift.  The result is intrinsically surprising: were the metal aversion to remain in effect for these objects, we would expect their occurrence (if still in the high metallicity range) to be far lower than the typical metallicity for the population at that luminosity and redshift (i.e., either a outlier of said population, or among the lowest galaxies available within it). While the majority of the LGRB population is constrained to low metallicities of about a third solar and below these exceptions probably show that is it still possible to still form an LGRB in a high metallicity environment although with greater rarity. 

From this we conclude that there are three possible explanations for the presence of the LGRBs observed in high metallicity hosts as seen to date:  (1) LGRBs do not occur in high metallicity environments and those seen in high metallicity hosts are in fact occurring in low metallicity environments that have become associated with otherwise high metallicity hosts but remain unenriched. (2) The LGRB formation mechanism while preferring low metallicity environments does not strictly require it resulting in a gradual decline in burst formation with increasing metallicity. (3) The typical low metallicity LGRBs and the few high metallicity cases are the result of physically different burst formation pathways with only the former affected by the metallicity and the later occurring much more infrequently.  To discriminate between these possibilities we recommend scrutiny of the metallicity distribution of high metallicity LGRBs within their host galaxies (i.e. Do they favor low metallicity regions, or do they track the brightest stars, as seen in the general LGRB population \citep{Fruchter}).

\section{Conclusions}

Here we present our imaging and spectroscopic observations of the host galaxies of two dark long bursts with anomalously high metallicities, LGRB 051022 and LGRB 020819B, which, in conjunction with another LGRB event with an optical afterglow \citep{Levesque2}, comprise the known three LGRBs with high metallicity host galaxies.  In \cite{stats_paper} we analyze the metallicity distribution of the LGRB population at large as well as comparing and contrasting with the general star-forming galaxies and supernovae populations to conclude that the LGRB hosts are significantly depressed in metallically due to an intrinsic metal aversion preference in LGRB formation.  Here, we focused on the few exceptions to this trend: high metallicity LGRB events, whose occurrence is astonishingly rare compared to the much greater volume of star-formation available at high metallicity.

Most notably, aside form their existence, is that these high metallicity LGRBs lack any apparent preference for a low metallically environment either with regard to other galaxies of similar luminosity and redshift or the location of the burst occurrence within their host galaxies.  We thus conclude that despite a massive preference for low metallicity in LGRB formation, once that threshold in grossly exceeded, there remains no marginal preference for a lower metallically.  This result is intrinsically surprising, as were the metal aversion effect to remain in effect for these objects we would expect their occurrence, if still in the high metallicity range, to be far lower than the typical metallicity for the population at that luminosity and redshift (i.e. either a outlier of said population or among the lowest galaxies available within it).

We do however find this result to be consistent with that of another paper we are publishing concurrently.  In \cite{diff_rate_letter} we extended the \cite{stats_paper} analysis by normalizing the LGRB rate to the rate of underlying star-formation across different metallicities to directly probe and quantize how much more likely is an LGRB to form at one metallicity as compared with another.  We find that the gradient in LGRBs per unit star-formation is comparably flat at high metallicities after undergoing a sharp decline at log(O/H)+12 $\sim$ 8.3.  Understanding this metallicity cutoff is essential to efforts to compare the LGRB rate with that of star-formation as a function of redshift \citep{form_rate_letter}.

From this we conclude that there are three possible explanations for the presence of the LGRBs observed in high metallicity hosts as seen to date:  (1) LGRBs do not occur in high metallicity environments and those seen in high metallicity hosts are in fact occurring in low metallicity environments that have become associated with otherwise high metallicity hosts but remain unenriched. (2) The LGRB formation mechanism while preferring low metallicity environments does not strictly require it resulting in a gradual decline in burst formation with increasing metallicity. (3) The typical low metallicity LGRBs and the few high metallicity cases are the result of physically different burst formation pathways with only the former affected by the metallicity and the later occurring much more infrequently.

To discriminate between these possibilities we recommend scrutiny of the metallicity distribution of high metallicity LGRBs within their host galaxies (i.e. Do they favor low metallicity regions, or do they track the brightest stars, as seen in the general LGRB population a la \citealt{Fruchter}).  Such spatially resolved host spectroscopy have been conducted via IFU survey on one of the closest bursts, LGRB 980425 \citep{Christensen980425} and could be extended to slightly higher redshifts (especially for brighter \& bigger galaxies).  For the higher redshift targets z $\gtrsim$ 0.8 we propose a combination of laser guide star adaptive optics infrared integral field unit spectroscopy and tunable narrow band ACS ramp filter observations to overcome the limits of ground based seeing.

Finally we hypothesize on binaries as the possible source of a second lesser traveled and metallicity independent LGRB formation pathway.  Assuming that the specific star-forming regions hosting the bursts and the burst progenitors themselves are high metallicity systems this significantly complicates current theories of LGRB formation.  These theories nearly universally require a rapidly spinning progenitor to collimate the formation of jets required for GRB emission.  The higher mass loss rates of high metallicity stars would thus bleed off the requisite angular momentum and prevent the formation of jets capable of escaping their host star.  The presence of a binary companion could, under the right circumstances, spin up the progenitor star allowing for the jet formation even in high metallicity environments.

\acknowledgments

Based on observations obtained at the Gemini Observatory acquired through the Gemini Science Archive and processed using the Gemini IRAF package, which is operated by the Association of Universities for Research in Astronomy, Inc., under a cooperative agreement with the NSF on behalf of the Gemini partnership: the National Science Foundation (United States), the National Research Council (Canada), CONICYT (Chile), the Australian Research Council (Australia), MinistŽrio da Cincia, Tecnologia e Inova‹o (Brazil) and Ministerio de Ciencia, Tecnolog'a e Innovaci—n Productiva (Argentina).

The W.M. Keck Observatory is operated as a scientific partnership among the California Institute of Technology, the University of California and the National Aeronautics and Space Administration.  The  Observatory was made possible by the generous financial support of the  W.M. Keck Foundation.  The authors wish to recognize and acknowledge the very significant cultural role and reverence that the summit of Mauna Kea has always had within the indigenous Hawaiian community.  We are most fortunate to have the opportunity to conduct observations from this mountain.

Based on observations made with the NASA/ESA Hubble Space Telescope, obtained from the Data Archive at the Space Telescope Science Institute, which is operated by the Association of Universities for Research in Astronomy, Inc., under NASA contract NAS 5-26555. These observations are associated with program $\#$ 11343.

Support for this work was provided by NASA through grant number $\_\_\_\_\_\_\_$ from the Space Telescope Science Institute, which is operated by AURA, Inc., under NASA contract NAS 5-26555.

{\it Facilities:} \facility{Gemini:Gillett (GMOS)}, \facility{Keck:II (NIRSPEC)}, \facility{HST (ACS/WFC, WFC3/IR)}, \facility{MDM:Hiltner (RETROCAM)}

\bibliographystyle{apj_links}
\bibliography{\jobname}

\end{document}